\documentclass[%
nofootinbib,
 amsmath,amssymb,
 aps,
showkeys,
]{revtex4-2}

\usepackage{graphicx}
\usepackage{dcolumn}
\usepackage{bm}
\usepackage{subfigure}
\usepackage[mathlines]{lineno}
\usepackage{hyperref}
\hypersetup{colorlinks=true
  ,urlcolor=blue
  ,anchorcolor=blue
  ,citecolor=blue
  ,filecolor=blue
  ,linkcolor=blue
  ,menucolor=blue
  ,linktocpage=true
}

\begin{document}


\title{Quadratic Rastall Gravity: From Low-mass HESS J1731$-$347 to High-mass PSR J0952$-$0607 Pulsars}

\author{Waleed El~Hanafy}
 \affiliation{Centre for Theoretical Physics, The British University in Egypt, P.O. Box 43, El Sherouk City, Cairo 11837, Egypt}
 \email{waleed.elhanafy@bue.edu.eg}

\graphicspath{{./}{Figs/}}


\begin{abstract}
Similar to Rastall gravity we introduce matter-geometry nonminimal coupling which is proportional to the gradient of quadratic curvature invariants. Those are mimicking the conformal trace anomaly when backreaction of the quantum fields to a curved spacetime geometry is considered. We consider a static spherically symmetric stellar structure with anisotropic fluid and Krori-Barua metric potentials model to examine the theory. Confronting the model with NICER+XMM-Newton observational constraints on the pulsar PSR J0740$+$6620 quantifies the amount of the nonminimal coupling via a dimensionless parameter $\epsilon\simeq -0.01$. We verify that the conformal symmetry is broken everywhere inside the pulsar as the trace anomaly $\Delta>0$, or equivalently the trace of the stress-energy tensor $\mathfrak{T}<0$, whereas the adiabatic sound speed does not violate the conjecture conformal upper limit $v_r^2/c^2 = 1/3$. The maximum compactness accordingly is $C_\text{max}=0.752$ which is $4\%$ higher than GR. Notably, if the conformal sound speed constraint is hold, observational data excludes $\epsilon \geq 0$ up to $\geq 1.6\sigma$. The stellar model is consistent with the self-bound structure with soft linear equation of state. Investigating possible connection with MIT bag model of strange quarks sets physical bounds from microscopic physics which confirm the negative value of the parameter $\epsilon$. We estimate a radius $R=13.21 \pm 0.96$ km of the most massive observed compact star PSR J0952$-$0607 with $M=2.35\pm0.17 M_\odot$. Finally, we show that the corresponding mass-radius diagram fits well lowest-mass pulsar HESS J1731$-$347 and highest-mass pulsar PSR J0952$-$0607 ever observed as well as the intermediate mass range as obtained by NICER and LIGO/Virgo observations.
\end{abstract}

\maketitle


\section{Introduction}\label{Sec:intro}


The \textit{principle of minimal coupling} has been assumed by the general relativity (GR) theory which leads to a divergence-free energy-momentum constraint, $\nabla{_\alpha}\mathfrak{T}{^\alpha}{_\beta}=0$, in a curved spacetime. However, a more general framework has been argued by Rastall by giving up the minimal coupling principle and considering matter-geometry nonminimal coupling instead \cite{Rastall:1972swe,Rastall:1976uh}. In Rastall's gravity, the nonminimal coupling is assumed to be of linear order of Ricci scalar $\mathfrak{R}$, i.e. $\nabla{_\alpha}\mathfrak{T}{^\alpha}{_\beta} \propto \partial_\beta \mathfrak{R}$. Clearly, the theory becomes distinguishable from GR where high curvature associated with dense matter shows up, while it reduces to GR in vacuum. In this sense, compact stars such as in the case of neutron stars (NSs) are ideal regimes to test Rastall gravity, which has been shown to have novel features when applied to compact stellar objects \cite{ElHanafy:2022kjl,ElHanafy:2023vig}. On the contrary, it has been argued that Rastall gravity is completely equivalent to GR \cite{Visser:2017gpz}. However, this argument has been criticized and shown to be wrong \cite{Darabi:2017coc} due to misinterpretation of the matter stress-energy tensor. In practice, the inequivalence of the two theories has been ensured in several studies on thermodynamical, cosmological and astrophysical levels, c.f. \cite{Lin:2020fue,Li:2019jkv,2020EPJP..135..916Z,Oliveira:2015lka,Moradpour:2016fur,Bamba:2018zil,Cruz:2019jiq,Fabris:2020uey,daSilva:2020okh,Li:2022wzi,Nashed:2022zyi}.

Trace anomaly, on the other hand, is important when the backreaction of the quantum fields to a curved spacetime geometry is considered. It is well known that the quantum corrections due to trace anomaly are geometric in nature and can be expressed in terms of Gauss-Bonnet four dimensional topological invariant (Type A) and the square of the Weyl tensor (Type B) \cite{Duff:1993wm}. Those are, in general, of quadratic order of curvature invariants. At high densities, $n\sim 40 n_\text{nuc}$ where $n_\text{nuc}\approx 0.16$ fm$^{-3}$ is the baryon density number at nuclear saturation, perturbative quantum chromodynamics (pQCD) is applied, conformal symmetry is hold, i.e. the trace anomaly vanishes. At finite densities and/or temperatures compatible with NS densities, few times nuclear saturation density $\rho_\text{nuc}\sim 2.7\times 10^{14}$ g/cm$^{3}$, it has been argued that the conformal symmetry would be broken, i.e. trace anomaly does not vanish, due to running of the strong coupling with energy at quantum level \cite{Fujimoto:2022ohj}. Thermodynamic stability and causality indeed set a constraining physical range to the trace anomaly, $\Delta \in [1/3,-2/3]$, but they cannot determine its sign inside an NS.

It is believed that quantum phenomena should play an essential role in gravitational collapse of massive objects. On the classical level, the breaking of minimal coupling provides an approach to this problem \cite{Goenner:1976}. \textit{In the present work, we extend the nonminimal coupling frame to include exactly quadratic curvature invariants, which we call quadratic Rastall gravity (QRG). Since quadratic nonminimal coupling is motivated by quantum phenomena, we argue the possibility of combining these two approaches in a single frame by considering theories such as QRG}. We still have the privilege of recovering GR vacuum solutions, whereas quadratic contribution is more compatible with trace anomaly when the backreaction of the quantum fields to a curved spacetime geometry is considered. For sure, different set up is needed if one considers exactly trace anomaly. However, the present work may provide a step towards the exact problem.

For the classical matter sector, it is unlikely for a high dense matter as in the NS core to be represented by the ideally isotropic fluid. Even in low density regimes on the Newtonian limit, local anisotropy in matter fields could arise due to anisotropic velocity distribution. At high density regimes, local anisotropy naturally appears due to solidification \cite{Palmer:1974hb}, pion condensation \cite{Sawyer:1972cq}, super-fluidity \cite{Ramanan:2019kwf}, strong magnetic fields \cite{Weber:2006ep}, hyperons \cite{Rahmansyah:2020gar} and strong interactions \cite{bowers1974anisotropic}. \textit{Therefore, we assume the fluid to be anisotropic which is the more realistic case in the compact stars}. Local anisotropy and its consequences on the stability and the gravitational collapse of self-gravitating systems has been investigated by Herrera and Santos \cite{herrera1997local} (see also \cite{herrera1992cracking,Herrera:2007kz,Herrera:2008bt,Herrera:2011cr}). Several studies assume gravitationally bound (hadrons) or self bound (quark) stars according to the matter assumed. \textit{In the present study, we do not assume a specific matter type in prior, since no particular equation of state (EoS) is assumed. Instead, we assume the metric potentials inside the compact star to follow Krori-Barua (KB) ansatz \cite{Krori1975ASS}, which effectively relates pressure and density}.

NS is a core of collapsed massive star with unique characteristics: Mass $M=1.4 M_\odot$ and radius $R=10$ km for canonical NS, rapid rotation 1.39 ms and intense magnetic field $10^{12}$ Gauss. This makes the NS an astrophysical lighthouse where radio beams are emitted from its magnetic poles and received as pulses due to its spin. For this reason, it is called as a pulsar and in the case of rapid spin with ms period it is called Millisecond Pulsars (MSPs). It is widely believed that pulsars are in fact neutron stars while others suggest quark star model \cite{Bhattacharyya:2016kte,Annala:2019puf}. As a matter of fact, pulsars do not only provide a perfect laboratory to examine the physical properties of matter at high densities and compactness at Black Hole (BH) lower limits, but also to prob possible modified gravity scenarios in general and nonminimal coupling ones in particular. Fortunately, the availability of pulsar data using different multi-messenger astrophysics \cite{Meszaros:2019xej} with unprecedented precision provides an extremely important tool to constrain the parameter space of a suggested modified gravity.

Radio pulse times of arrival (ToAs) from pulsars are being used for decades to measure pulsars' masses via relativistic Shapiro time delay, while measuring the pulsar radius is relatively not an easy task. Neutron Star Interior Composition Explorer (NICER) is devoted to measure pulsars radii by tracing hot-spot regions on the pulsar surface which emit X-rays \cite{2017NatAs...1..895G}. NICER measurements of the radii of the pulsars PSR J0030+0451 and PSR J0740+6620 represent a challenge to theoretical models, since their masses are respectively $M=1.44^{+0.15}_{-0.14} M_\odot$ and $M= 2.08 \pm 0.07 M_\odot$ \citep{NANOGrav:2019jur,Fonseca:2021wxt} while their radii are almost $R\sim 13$ km. More precisely NICER measures the pulsar PSR J0030+0451 radius $R= 13.02_{-1.06}^{+1.24}$ km \cite{Miller:2019cac} and another independent measurement $R= 12.71^{+1.14}_{-1.19}$ km \cite{Raaijmakers:2019qny}, while the analysis of NICER + X-ray Multi-Mirror (XMM) Newton data of the pulsar PSR J0740+6620 measures $R= 13.7_{-1.5}^{+2.6}$ km \citep{Miller:2021qha} and $R=12.39_{-0.98}^{+1.30}$ km (68\% credible level) \cite{Riley:2021pdl}. In this sense, the more squeezable models, which suggest the neutrons to squeeze to quarks, seem to be disfavored by NICER observations. Notably, the latest analysis of PSR J0030+0451 using an updated NICER + XMM-Newton data (using ST+PDT model) measures $R=11.71^{+0.88}_{-0.83}$ \citep{Vinciguerra:2023qxq}. If the latest is confirmed, quark star model could be no longer disfavored. Nevertheless, massive pulsars $M \simeq 2 M_\odot$ such as PSR J0740+6620 represents a challenge to the less squeezable models, since the sound speed should rapidly increase somewhere inside the stellar core to large values $c_s^2\sim 0.7 c^2$. This requires a problematic nonmonotonical behaviour of the sound speed violating the conjectured conformal upper bound $c_s=c/\sqrt{3}$ \citep{McLerran:2018hbz,Altiparmak:2022bke}. It is noted that existence of massive pulsars points out a stiff EoS matter where the sound speed to speed of light ratio becomes large. On the other hand, gravitational wave signals, GW170817 \citep{TheLIGOScientific:2017qsa,LIGOScientific:2018cki} and GW190425 \citep{LIGOScientific:2020aai}, indicate no tidal deformability of the measured signals, as confirmed by Laser
Interferometer Gravitational-Wave Observatory (LIGO) and Virgo collaboration, which points out a soft EoS matter where the speed of sound is not large. It is to be noted that soft EoS is compatible with the conjectured conformal upper limit on the sound speed which evidently favors quark star models. On the other extreme, observations of the low-mass pulsar in supernova remnant HESS J1731\textendash{347}, with mass $M=0.77_{-0.17}^{+0.20}$ and radius $R=10.4^{+0.86}_{-0.78}$, suggest that the pulsar could be a light strange quark star \citep{2022NatAs...6.1444D,Horvath:2023uwl}.

We present the organization of the paper in some detail as follows: In Sec. \ref{Sec:QRgrav}, we derive the field equations of QRG. In Sec. \ref{Sec:SSS_mod}, we set up the stellar model assuming that the spacetime is described by Static Spherically Symmetric (SSS) configuration, the metric potentials are given by KB ansatz and the matter fluid is anisotropic. We also investigate corresponding matter EoS which has been shown to have a linear pattern and matching conditions which determine the model parameters. In Sec. \ref{Sec:constr}, we utilize the astrophysical constraints on the mass and radius of the pulsar PSR J0740+6620 from NICER+XMM-Newton data to determine the numerical value of the nonminimal coupling parameter $\epsilon$ as well as the KB model parameters. Consequently, we determine the radial dependence of the sound speed and the conformal anomaly inside the pulsar. In addition, we investigate different constitutional forces which are in hydrostatic equilibrium including the newly introduced one due to quadratic nonminimal coupling. Similarly, we obtain the radial dependence of the adiabatic index which helps to distinguish gravitational bound star from self bound one. Our calculations show that the pressures and densities fit well with linear model which confirms the linear pattern of the KB induced EoS. In Sec. \ref{Sec:features}, we discuss some physical features of the obtained model showing that the observational data excludes $\epsilon\geq 0$, which may indicate some underlying similarity between QRG and conformal anomaly when quantum fields backreaction is included. We use the Trace Energy Condition (TEC) to determine the maximum possible compactness $C=2GMR^{-1}c^{-2}$. Moreover, we investigate in details possible connection between QRG and MIT bag model of strange quark star which sets further constraints on the model parameters but from microscopic structure of the star. In Sec. \ref{Sec:J0952–0607}, by assuming that the conformal constraints on the sound speed to be hold at the core of the pulsar PSR J0952–0607, we estimate the pulsar radius accordingly. We also provide mass-radius diagram which confirms the validity of the model to describe compact stars from lightest-to-heaviest pulsars ever observed including recent constraints on the intermediate mass range from NICER and LIGO/Virgo collaboration. In Sec. \ref{Sec:summary}, we summarize and conclude the work. Furthermore, we followed the last section by two appendices \ref{App:WFL} and \ref{App:pulsars} to discuss weak field limit of the field equations and to confront the model with more pulsars’ observational data of different observational techniques.


\section{Quadratic Rastall Gravity}\label{Sec:QRgrav}


We briefly review the main structure and assumptions which lead to GR and Rastall field equations. Then, we extend the discussion in details to the QRG. It is well known that the double contraction of Bianchi identity in Riemannian geometry leads to the divergence-free second order tensor,
\begin{equation}\label{eq:Bianchi}
    \nabla{_\alpha}\mathfrak{G}{^\alpha}{_\beta}=  \nabla{_\alpha}(\mathfrak{R}{^\alpha}{_\beta}-\frac{1}{2}\delta^\alpha_\beta\mathfrak{R})\equiv 0,
\end{equation}
where $\nabla_\alpha$ denotes the covariant derivative associated with Levi-Civita linear connection, the second order tensors $\mathfrak{G}{^\alpha}{_\beta}$ and $\mathfrak{R}{^\alpha}{_\beta}$ denote Einstein and Ricci tensors respectively, whereas the contraction of the latter obtains Ricci scalar $\mathfrak{R}=g^{\alpha\beta}\mathfrak{R}_{\alpha\beta}$. On the other hand, the minimal coupling principle (divergence-free stress-energy tensor) has been introduced to generalize its validity in special relativity to curved spacetimes. Then, it yields
\begin{equation}\label{eq:MCP}
    \nabla_{\alpha}\mathfrak{T}{^\alpha}{_\beta}=0.
\end{equation}
This finally led to write the GR field equations
\begin{equation}\label{eq:GR}
    \mathfrak{G}{^\alpha}{_\beta}=\kappa_E \mathfrak{T}{^\alpha}{_\beta},
\end{equation}
where $\kappa_E$ denotes Einstein coupling constant, which is given in terms of $G$ is the Newtonian gravitational constant $G$ and the speed of light $c$ by $\kappa_E=8\pi G/c^4$. The left hand side represents the geometric sector configuring the spacetime and the right hand side represents the matter sector whereas matter-geometry nonminimal coupling vanishes.

In Rastall's gravity, the minimal coupling principle \eqref{eq:MCP} is no longer assumed, and replaced by \citep{Rastall:1972swe,Rastall:1976uh}
\begin{equation}\label{eq:Rastall}
     \nabla_{\alpha}\mathfrak{T}{^\alpha}{_\beta}=\varepsilon\, \partial_{\beta} \mathfrak{R}.
\end{equation}
We use $\varepsilon\to -\frac{\varepsilon}{\kappa_R}$ where the negative sign is a convention, $\varepsilon$ is dimensionless and $\kappa_R$ is Rastall's coupling constant. This leads to a modified version of Einstein field equations which account for matter-geometry nonminimal coupling corresponds to Rastall assumption \eqref{eq:Rastall} as below
\begin{equation}\label{eq:RT}
    \mathfrak{G}{^\alpha}{_\beta}=\mathfrak{R}{^\alpha}{_\beta}-\frac{1}{2}\delta^\alpha_\beta\mathfrak{R}=\kappa_R(\mathfrak{T}{^\alpha}{_\beta}+\frac{\varepsilon}{\kappa_R} \delta^\alpha_\beta\mathfrak{R}).
\end{equation}
The contraction of the above field equation,
\begin{equation*}
    \mathfrak{R}=-\frac{\kappa_R}{1+4\epsilon}\mathfrak{T},
\end{equation*}
puts a restriction $\epsilon\neq -\frac{1}{4}$. The weak field limit of the above field equations yields \citep[c.f.][]{Rastall:1972swe,Moradpour:2017ycq,Moradpour:2016ubd}
\begin{equation}\label{eq:Rconst}
    \kappa_R=\eta\,\kappa_E \text{ and } \eta=\frac{1+4\varepsilon}{1+6\varepsilon}, \qquad (\varepsilon\neq -\frac{1}{6})
\end{equation}
where $\kappa_R$ has the same dimension of $\kappa_E$, whereas $\kappa_R=\kappa_E$ if and only if $\eta=1$ (equivalently $\varepsilon=0$).

In the present paper, we take the matter-geometry nonminimal coupling to be proportional to the gradient of quadratic curvature invariants as follows
\begin{equation}\label{eq:QRastall}
     \nabla_{\alpha}\mathfrak{T}{^\alpha}{_\beta}=-\frac{\epsilon \ell^2}{\kappa}\, \partial_{\beta} \left(\mathfrak{R}^2-\mathfrak{R}_{\mu \nu}\mathfrak{R}^{\mu \nu}\right),
\end{equation}
where $\epsilon$ is a dimensionless parameter, $\ell$ is a constant of dimension [L] characterizing the mass of the matter distribution, i.e. it cannot be fitted arbitrarily to observational data\footnote{In astrophysical applications we replace $\ell$ by the star radius $R$ which is a reasonable choice.}, and $\kappa$ acquires the same dimension as $\kappa_E$. Thus, the nonminimal coupling is characterized by the dimensionless $\epsilon$-parameter, where the corresponding modified field equations are
\begin{equation}\label{eq:QRT}
    \mathfrak{G}{^\alpha}{_\beta}=\kappa\left[\mathfrak{T}{^\alpha}{_\beta}+\frac{\epsilon \ell^2}{\kappa} \delta{^\alpha_\beta}\left(\mathfrak{R}^2-\mathfrak{R}_{\mu \nu}\mathfrak{R}^{\mu \nu}\right)\right].
\end{equation}
The above field equations will be referred to as QRG. Contracting the above equation gives rise to
\begin{equation}
    -\mathfrak{R}=\kappa\left[\mathfrak{T}+\frac{4\epsilon \ell^2}{\kappa}\left(\mathfrak{R}^2-\mathfrak{R}_{\mu \nu}\mathfrak{R}^{\mu \nu}\right)\right],
\end{equation}
where $\mathfrak{T}:=g_{\alpha\beta}\mathfrak{T}^{\alpha\beta}$ is the trace of the matter stress-energy tensor. For empty spacetime, i.e. $\mathfrak{T}_{\alpha\beta}=0$, we obtain
\begin{equation}
    \left(1+4\epsilon \ell^2 \mathfrak{R}\right) g^{\mu \nu}\mathfrak{R}_{\mu \nu}=4\epsilon \ell^2 \mathfrak{R}^{\mu\nu}\mathfrak{R}_{\mu\nu}.
\end{equation}
Clearly $\mathfrak{R}_{\mu\nu}=0$ is a solution of the above constraint, then GR vacuum solutions are recovered in QRG. More interestingly, for nondegenerate $\mathfrak{R}_{\mu\nu}$, another vacuum state can be obtained as $\mathfrak{R}=\frac{-1}{3\epsilon \ell^2}$, where $\epsilon \neq 0$, which gives rise to de Sitter/Anti-de Sitter vacuum according to the sign of $\epsilon$. The latter vacuum case does not allow recovering GR as $\epsilon=0$ is not valid and we drop this case in the present study. Therefore, we still have the privilege of recovering GR vacuum solutions, as in the Rastall gravity, where $\mathfrak{R}_{\mu\nu}=0=\mathfrak{R}$ in empty spacetime. However, the QRG includes only quadratic terms of curvature invariants and the Newtonian limit of the present theory is expected to be identical to the Newtonian limit of GR unlike RT, i.e. $\kappa=\kappa_E=8\pi G/c^4$, see appendix \ref{App:WFL}. Therefore, all solar system tests are fulfilled in QRG. Another motivation to consider quadratic corrections is the trace anomaly when the backreaction of the quantum fields to a curved spacetime geometry is considered. It is well known that the quantum corrections due to trace anomaly are geometric in nature and can be expressed in terms of Gauss-Bonnet four dimensional topological invariant (Type A) and the square of the Weyl tensor (Type B). Therefore, it is reasonable to consider quadratic corrections of curvature invariants for matter-geometry nonminimal coupling as in equations \eqref{eq:QRT}. Notably the sign of $\epsilon$-like parameter of the trace anomaly must be negative, it however remains undetermined in our approach at this stage to be discussed in Subsec. \ref{Sec:neg_para}.

It is convenient to rewrite the field equations \eqref{eq:QRT} in the following form
\begin{equation}\label{eq:eff_QRG}
    \mathfrak{G}_{\alpha\beta}=\mathfrak{R}_{\alpha\beta}-\frac{1}{2}g_{\alpha\beta}\mathfrak{R}=\kappa \widetilde{\mathfrak{T}}_{\alpha\beta},
\end{equation}
where the tensor $\widetilde{\mathfrak{T}}_{\alpha\beta}$ represents the total stress-energy tensor
\begin{equation}\label{eq:eff_Tab}
\widetilde{\mathfrak{T}}_{\alpha\beta}:=\mathfrak{T}_{\alpha\beta}+\frac{\epsilon \ell^2}{\kappa} g_{\alpha\beta}\left(\mathfrak{R}^2-\mathfrak{R}_{\mu \nu}\mathfrak{R}^{\mu \nu}\right).
\end{equation}
By virtue of \eqref{eq:QRastall}, it is obvious that $\widetilde{\mathfrak{T}}_{\alpha\beta}$ a divergence-free, i.e.
\begin{equation}
    \nabla_{\alpha}\widetilde{\mathfrak{T}}{^\alpha}{_\beta}=0.
\end{equation}
It can be shown that \eqref{eq:eff_QRG} has an equivalent form
\begin{equation}\label{eq:Ricci_tensor}
    \mathfrak{R}_{\alpha\beta}=\kappa\left(\widetilde{\mathfrak{T}}_{\alpha\beta}-\frac{1}{2}g_{\alpha\beta}\widetilde{\mathfrak{T}}\right),
\end{equation}
where $\widetilde{\mathfrak{T}}:=g_{\mu\nu}\widetilde{\mathfrak{T}}^{\mu\nu}$ represents the trace of the effective stress-energy tensor. Consequently, we write the following important relation
\begin{equation}
    \mathfrak{R}^2-\mathfrak{R}_{\mu\nu}\mathfrak{R}^{\mu\nu}=\kappa^2\left(\widetilde{\mathfrak{T}}{^2}-\widetilde{\mathfrak{T}}_{\mu\nu}\widetilde{\mathfrak{T}}^{\mu\nu}\right).
\end{equation}
Substituting into \eqref{eq:eff_Tab}, we obtain
\begin{equation}\label{eq:QRT2}
    \mathfrak{T}_{\alpha\beta}=\widetilde{\mathfrak{T}}_{\alpha\beta}-\epsilon \kappa \ell^2 g_{\alpha\beta}\left(\widetilde{\mathfrak{T}}{^2}-\widetilde{\mathfrak{T}}_{\mu\nu}\widetilde{\mathfrak{T}}^{\mu\nu}\right).
\end{equation}
The contracted form of the above equations yields
\begin{equation}
    \mathfrak{T}=\widetilde{\mathfrak{T}}-4\epsilon \kappa \ell^2 \left(\widetilde{\mathfrak{T}}{^2}-\widetilde{\mathfrak{T}}_{\mu\nu}\widetilde{\mathfrak{T}}^{\mu\nu}\right).
\end{equation}
To summarize, the QRG field equations are given by \eqref{eq:QRT}, or equivalently by \eqref{eq:QRT2} where the effective energy-stress tensor on the right hand side is determined by the geometric sector, i.e. Einstein tensor, as given by \eqref{eq:eff_QRG}. The latter choice of the QRG field equations, namely \eqref{eq:QRT2}, obtains the matter sector in the left hand side and the geometric sector in the right hand side. This form clearly describes how the matter sector, in nonempty spacetime, would behave in QRG distinguishable from GR as long as $\epsilon\neq 0$. On another word, $\mathfrak{T}_{\alpha\beta}=\widetilde{\mathfrak{T}}_{\alpha\beta}$ if and only if $\epsilon=0$. One natural laboratory to test the present proposal is compact stars where dense matter and high curvature are expected to manifest matter-geometry nonminimal coupling scenario. Fortunately, recent precise measurements of mass and radius of compact stars via several astrophysical observations could set strict bounds on the nonminimal coupling parameter $\epsilon$.


\section{Static Spherically Symmetric Model}\label{Sec:SSS_mod}


In this study, we assume that the stellar structure is static spherically symmetric in 4-dimensional spacetime, where the coordinates are spherical polar ($0 \leq t < \infty,0\leq r < \infty, 0 \leq \theta < 2\pi,0 \leq \phi \leq \pi$). This gives rise to the line element
\begin{equation}\label{eq:metric}
    ds^2=-e^{\alpha}c^2 dt^2 + e^{\beta} dr^2+ r^2 (d\theta^2+\sin^2 \theta \, d\phi^2),
\end{equation}
where $\alpha\equiv \alpha(r)$ and $\beta\equiv \beta(r)$ are the metric function with only radial dependence. Since we aim to test the present theory in high dense medium, it is reasonable to assume anisotropic fluid case as discussed in the introduction, where the stress-energy tensor in this case in its standard form reads\footnote{We use signature convention $(-,+,+,+)$.} \cite{herrera1997local}
\begin{equation}\label{Tmn-anisotropy}
    \mathfrak{T}{^\alpha}{_\beta}=(\rho c^2+p_{t})v{^\alpha} v{_\beta}+p_{t} \delta ^\alpha _\beta + (p_{r}-p_{t}) w{^\alpha} w{_\beta},
\end{equation}
where
\begin{equation*}
    v^\alpha v_\alpha=-c^2,\, w^\alpha w_\alpha=1,\, v^\alpha w_\alpha=0,
\end{equation*}
and the four-velocity (time-like) vector field $v^\alpha=(c e^{-\alpha/2},0,0,0)$, the unit vector field(space-like) in the radial direction $w{^\alpha}=(0,e^{-\beta/2},0,0,0)$. In this case, the component of the energy-stress tensor $\mathfrak{T}{^0}{_0}=-\rho(r) c^2$ represents the fluid density, $\mathfrak{T}{^1}{_1}=p_r(r)$ represents the radial pressure ($w_\parallel$) and $\mathfrak{T}{^2}{_2}=\mathfrak{T}{^3}{_3}=p_t(r)$ represent tangential pressures ($w_\perp$), i.e. the matter tensor reduces to the diagonal form $\mathfrak{T}{^\alpha}{_\beta}=diag(-\rho c^2,\,p_{r},\,p_{t},\,p_{t})$. Similarly, the effective stress-energy tensor can be given by $\widetilde{\mathfrak{T}}{^\alpha}{_\beta}=diag(-\tilde{\rho} c^2,\,\tilde{p}_{r},\,\tilde{p}_{t},\,\tilde{p}_{t})$. Applying QRG field equations \eqref{eq:QRT2} to the spacetime \eqref{eq:metric} where the matter sector is as given by \eqref{Tmn-anisotropy}, we obtain the following
\begin{eqnarray}
\rho c^2&=&\tilde{\rho} c^2-2\epsilon \kappa \ell^2 \left[\tilde{\rho} c^2 (\tilde{p}_r+2\tilde{p}_t)-\tilde{p}_t(\tilde{p}_t+2\tilde{p}_r)\right],\nonumber\\[5pt]
p_r&=&\tilde{p}_r+2\epsilon \kappa \ell^2 \left[\tilde{\rho} c^2 (\tilde{p}_r+2\tilde{p}_t)-\tilde{p}_t(\tilde{p}_t+2\tilde{p}_r)\right],\nonumber\\[5pt]
p_t&=&\tilde{p}_t+2\epsilon \kappa \ell^2 \left[\tilde{\rho} c^2 (\tilde{p}_r+2\tilde{p}_t)-\tilde{p}_t(\tilde{p}_t+2\tilde{p}_r)\right].\qquad
\label{eq:Feqs}
\end{eqnarray}
One should note that the right hand sides of the above equations are completely determined by the geometric sector, i.e. $\tilde{\rho}c^2=\frac{1}{\kappa}\mathfrak{G}{_0}{^0}$, $\tilde{p}_r=\frac{1}{\kappa}\mathfrak{G}{_1}{^1}$ and $\tilde{p}_t=\frac{1}{\kappa}\mathfrak{G}{_2}{^2}=\frac{1}{\kappa}\mathfrak{G}{_3}{^3}$. We express the anisotropy contribution as the pressure difference, $\delta(r) = p_t-p_r$, where $\delta=0$ in isotropic case ($p_t=p_r$). Using the above equations, it yields
\begin{equation}\label{eq:Delta1}
\delta(r) = p_t-p_r=\tilde{p}_t-\tilde{p}_r=\frac{e^{-\beta}}{4\kappa r^2}\left[(2\alpha''-\alpha'\beta'+\alpha'^2)r^2-2(\alpha'+\beta')r+4(e^\beta-1)\right],
\end{equation}
where prime and double prime denote first and second derivatives with respect to the radial coordinate. We remark that the matter-geometry coupling does not contribute to the anisotropy once the spherically symmetric spacetime configuration is assumed. Therefore, deviations from GR due to matter-geometry coupling cannot be spoiled up with various anisotropic effects. For the $\epsilon=0$ case, the differential equations \eqref{eq:Feqs} coincide with Einstein field equations of an interior spherically symmetrical spacetime \citep[c.f.,][]{Roupas:2020mvs}. Alternatively, the field equations \eqref{eq:Feqs} can be rewritten as
\begin{eqnarray}
\nonumber \tilde{\rho} c^2&=&\frac{e^{-\beta}}{\kappa r^2}(e^\beta+\beta' r -1)=\rho c^2-\frac{1+2\epsilon\kappa\ell^2(3\rho c^2-p_r-2p_t)}{12\epsilon\kappa\ell^2}-\frac{4\epsilon\kappa\ell^2(p_r-p_t)^2-2(p_r+2p_t)}{3\left[2\epsilon\kappa\ell^2(3\rho c^2+p_r+2p_t)+1-\sqrt{X(r)}\right]},\nonumber\\[8pt]
\tilde{p}_r&=&\frac{e^{-\beta}}{\kappa r^2}(1-e^\beta+\alpha' r)=p_r+\frac{1}{4}\left(\rho c^2- p_r - 2 p_t\right)+\frac{1-\sqrt{X(r)}}{24\epsilon\kappa\ell^2},\nonumber\\[8pt]
\tilde{p}_t&=&\frac{e^{-\beta}}{4\kappa r}\left[(2\alpha''-\alpha' \beta'+\alpha'^2)r+2(\alpha'-\beta')\right]
=p_t+\frac{1}{4}\left(\rho c^2- p_r - 2 p_t\right)+\frac{1-\sqrt{X(r)}}{24\epsilon\kappa\ell^2},\qquad
\label{eq:Feqs2}
\end{eqnarray}
where
\begin{equation}
    X(r)=1+12\epsilon\kappa\ell^2(\rho c^2-p_r-2p_t)+36\epsilon^2\kappa^2\ell^4\left[\frac{4}{3}p_t(\rho c^2-p_r+p_t)+\frac{1}{3}\rho c^2(3\rho c^2+2p_r)+p_r^2\right].
\end{equation}
The above equations \eqref{eq:Feqs2} allow to write the geometric sector on the left hand side in terms of the matter sector including nonminimal coupling terms. It is straightforward to check that the effective density and pressures reduce to the matter ones when the nonminimal coupling contribution vanishes as $\epsilon\to 0$.
\subsection{Krori-Barua ansatz}\label{Sec:KB}
We assume that the metric potentials inside the stellar model are as defined by Krori-Barua ansatz \cite{Krori1975ASS}, i.e.
\begin{equation}\label{eq:KB}
    \alpha(r)=a_0 r^2/R^2+a_1,\,  \beta(r)=a_2 r^2/R^2,
\end{equation}
where $R$ denotes the star radius. The dimensionless KB model parameters \{$a_0, a_1, a_2$\} are determined by matching conditions on the stellar boundary. Although KB potentials have been used widely in the literature, c.f. \cite{Roupas:2020mvs,ElHanafy:2022kjl,ElHanafy:2023vig}, it is worth to mention that the final output depends on the gravitational theory. The KB ansatz could be understood as an alternative of an equation of state. It is mainly used to obtain a nonsingular interior behaviour within the stellar configuration. Since the nature of matter inside compact stars at high densities and pressures is not known, we take the other approach by assuming KB potentials instead of assuming equations of state. However, the main differences would be a consequence of the modified gravity sector which clearly differentiates the present work from \cite{ElHanafy:2022kjl}. Similarly, one can use same equation of state of the matter sector, i.e. radiation, cold dark matter, in cosmological applications while the consequences will be different according to the modified gravity theory under investigation. It proves convenient to write the field equations in dimensionless forms, therefore we introduce the dimensionless radius $0 \leq x=r/R \leq 1$ and the dimensionless quantities
\begin{equation}
    \bar{\rho}(r)=\frac{\rho(r)}{\rho_{\star}},\, \bar{p}_r(r)=\frac{p_r(r)}{\rho_{\star} c^2},\, \bar{p}_t(r)=\frac{p_t(r)}{\rho_{\star} c^2}\,\text{and}\, \bar{\delta}(r)=\frac{\delta(r)}{\rho_{\star} c^2},
\end{equation}
where $\rho_{\star}=\frac{1}{\kappa c^2 R^2}$ denotes a characteristic density. Additionally, it is reasonable to take the characteristic length $\ell=R$. Substituting into the field equations \eqref{eq:Feqs} yields
\begin{eqnarray}
\bar{\rho}&=& \frac{e^{-a_2 x^2}}{x^2}(e^{a_2 x^2}-1+2a_2 x^2) \nonumber\\
&-&\frac{2\epsilon}{x^4}\left[\left(-a_0^2(a_0-a_2)^2 x^8-8a_0 (a_0-a_2) (a_0-3a_2/4)x^6+(24a_0a_2-16a_0^2-5a_2^2)x^4\right.\right.\nonumber\\
&&\left.\left.-2(5a_0-3a_2)x^2-1\right)e^{-2a_2x^2}+2\left(1+2a_0(a_0-a_2)x^4+(5a_0-3a_2)x^2\right)e^{-a_2x^2}-1\right],\nonumber\\[8pt]
\bar{p}_r&=&\frac{e^{-a_2 x^2}}{x^2}(1-e^{a_2 x^2}+2a_0 x^2) \nonumber\\
&+&\frac{2\epsilon}{x^4}\left[\left(-a_0^2(a_0-a_2)^2 x^8-8a_0 (a_0-a_2) (a_0-3a_2/4)x^6+(24a_0a_2-16a_0^2-5a_2^2)x^4\right.\right.\nonumber\\
&&\left.\left.-2(5a_0-3a_2)x^2-1\right)e^{-2a_2x^2}+2\left(1+2a_0(a_0-a_2)x^4+(5a_0-3a_2)x^2\right)e^{-a_2x^2}-1\right],\nonumber\\[8pt]
\bar{p}_t&=& e^{-a_2 x^2}(2 a_0-a_2 +a_0 (a_0 - a_2) x^2) \nonumber\\
&+&\frac{2\epsilon}{x^4}\left[\left(-a_0^2(a_0-a_2)^2 x^8-8a_0 (a_0-a_2) (a_0-3a_2/4)x^6+(24a_0a_2-16a_0^2-5a_2^2)x^4\right.\right.\nonumber\\
&&\left.\left.-2(5a_0-3a_2)x^2-1\right)e^{-2a_2x^2}+2\left(1+2a_0(a_0-a_2)x^4+(5a_0-3a_2)x^2\right)e^{-a_2x^2}-1\right],\nonumber\\[8pt]
\label{eq:Feqs3}
\end{eqnarray}
and the anisotropy indicator \eqref{eq:Delta1} reads
\begin{equation}\label{eq:Delta2}
    \bar{\delta}=\frac{e^{-a_2 x^2}}{x^2}\left[e^{a_2 x^2}-1+a_0(a_0-a_2)x^4 -a_2 x^2\right].
\end{equation}
The mass function which obtains the matter mass content within a radius $r$ is given by
\begin{equation}
    \mathfrak{M}(r)=4 \pi \int_{\xi=0}^r \rho(\xi) \, \xi^2 d\xi.
\end{equation}
Recalling the density profile \eqref{eq:Feqs2}, we find
\begin{equation}\label{eq:Mass}
    \mathfrak{M}(x)=\frac{M}{C}\left[x(1-e^{-a_2 x^2})+\epsilon \, \zeta(x)\right].
\end{equation}
where the compactness parameter $C$ and the function $\zeta(x)$ are
\begin{eqnarray}
C&=&\frac{2GM}{c^2 R}, \nonumber\\
\zeta(x)&=& \tfrac{a_2^{-\frac{7}{2}}}{2x}\left\{\right.\left[\right.-\tfrac{1}{4}(a_0x^2(78+51a_0x^2+8a_0^2x^4)+16)a_2^{\frac{7}{2}}+\left(\right.(a_0^2x^4+\tfrac{11}{2}a_0x^2+\tfrac{103}{16})a_0^2 a_2^{\frac{5}{2}}\nonumber\\
&&+(6a_0x^2+a_0^2x^4+5)a_2^{\frac{9}{2}}+(\tfrac{5}{4}a_0^4x^2+\tfrac{33}{8}a_0^3)a_2^{\frac{3}{2}}+\tfrac{15\sqrt{a_2}a_0^4}{16}\left.\right)x^2\left.\right]e^{-2a_2x^2}\nonumber\\
&&-8a_2^{\frac{5}{2}}\left[a_2-(a_0-a_2)a_0x^2\right]e^{-a_2x^2}+\tfrac{\sqrt{2\pi}15x}{64}\text{erf}(\sqrt{2a_2}x)\left[\tfrac{112a_2^4}{3}-\tfrac{328a_2^3 a_0}{15} - \tfrac{22a_2 a_0^3}{5} -\tfrac{103a_0^2 a_2^2}{15}-a_0^4\right]\nonumber\\
&&+4\sqrt{\pi} x a_2^2 (a_0+5a_2) (a_0-a_2)\text{erf}(\sqrt{a_2}x)+4a_2^{\frac{7}{2}}
\left.\right\}.
\end{eqnarray}
Obviously, the mass function \eqref{eq:Mass} determines $\mathfrak{M}(x\to 0)=0$, and reduces to the GR version \citep{Roupas:2020mvs} when $\epsilon \to 0$ .
\subsection{Equation of state}\label{Sec:EoS}
We note that the field equations \eqref{eq:Feqs} contain five unknown functions: Three matter sector functions, namely the density, radial pressure and tangential one ($\rho,~p_r,~p_t$), on the other hand there are two geometric sector functions, namely the metric potentials ($\alpha,\beta$). Therefore, in principle, two additional constraints are required in order to close the system. Those constraints could be EoSs $p_r(\rho)$ and $p_t(\rho)$, we instead use the KB ansatz \eqref{eq:KB}. However, this geometric approach enables us to write the field equations \eqref{eq:Feqs2} as given by the following relations up to $O(x^8)$
\begin{equation}\label{eq:EoSx}
    \bar{p}_r(\bar{\rho})\approx c_1 \bar{\rho} + c_0, \quad \bar{p}_t(\bar{\rho})\approx \tilde{c}_1 \bar{\rho} + \tilde{c}_0,
\end{equation}
where
\begin{eqnarray}\label{eq:EoS_const}
    c_0&=&{64(a_0+a_2)\left[a_2^2/32+\left(a_0^3-15a_0^2 a_2/4+35a_0 a_2^2/8-a_2^3\right)\epsilon\right]\over\left(32a_0^3-168a_0^2 a_2+260a_0a_2^2-80a_2^3\right)\epsilon-5a_2^2},\nonumber\\
    c_1&=&-1-{4a_2(a_0+a_2)\over\left(32a_0^3-168a_0^2 a_2+260a_0a_2^2-80a_2^3\right)\epsilon-5a_2^2},\nonumber\\
    \tilde{c}_0&=&{-a_2^3+8a_0a_2^2-6a_0^2a_2+\left(16a_0^4-8a_2a_0^3-152a_2^2a_0^2+324a_2^3a_0-88a_2^4\right)\epsilon\over\left(32a_0^3-168a_0^2 a_2+260a_0a_2^2-80a_2^3\right)\epsilon-5a_2^2},\nonumber\\
    \tilde{c}_1&=&-1-{3(a_0+a_2)^2-5a_0^2\over\left(32a_0^3-168a_0^2 a_2+260a_0a_2^2-80a_2^3\right)\epsilon-5a_2^2}.
\end{eqnarray}
The above relations are of an extreme importance when confront the model with astrophysical observations.  For example, vanishing of the parameter $c_0$ indicates that the star would be gravitationally bound (radial pressure at the stellar surface vanishes as density approaches zero), while significant value of this parameter indicates that the star would be self-bound (radial pressure vanishes at significant surface density). This reflects possible composition of the internal structure of the star core. On the other hand, the parameter $c_1$ is strongly related to the radial speed of sound inside the star fluid, $c_s^2=dp_r/d\rho$, see Subsec. \ref{Sec:conformal}, which indicates how much soft/stiff its EoS. In this sense, stiff EoS with gravitationally bound star may point out presence of hadron matter, whereas soft EoS with self-bound star would indicate strange quark matter content. This will be discussed in details in Subsec. \ref{Sec:structure}. Fortunately, all parameters in \eqref{eq:EoS_const} are completely determined by the compactness of the star and the nonminimal coupling parameter $\epsilon$, which in turn distinguishes QRG from GR, as will be discussed in the following subsection.
\subsection{Matching conditions}\label{Sec:Match}
It is easy to verify that exterior vacuum region in QRG is exactly the GR Schwarzschild solution, since both theories are identical in vacuum ($\mathfrak{R}_{\alpha\beta}=0$). Therefore, the exterior vacuum solution is given as
\begin{equation}
    ds^2=-\left(1-\frac{2GM}{c^2 r}\right) c^2 dt^2+\left(1-\frac{2GM}{c^2 r}\right)^{-1} dr^2+r^2 (d\theta^2+\sin^2 \theta d\phi^2).
\end{equation}
Recalling the interior spacetime \eqref{eq:metric}, utilizing KB ansatz \eqref{eq:KB}, we therefore have the boundary conditions (at $r=R$)
\begin{equation}\label{eq:bo}
    a_0+a_1=\ln(1-C)\,\text{and}\, a_2=-\ln(1-C).
\end{equation}
In addition to the vanishing of the radial pressure \eqref{eq:Feqs2} on the star boundary, i.e.
\begin{eqnarray}\label{eq:bo2}
    \bar{p}_r(x=1)&=&a_0+\frac{1}{2}(1-e^{a_2})+\epsilon\left[\right.e^{a_2}+2(2a_2-5)a_0+2(3a_2-2a_0^2)-2+e^{-a_2}\left\{\right.a_0^4-2\times\nonumber\\
    &&(a_2-4)a_0^3+(16-14a_2+a_2^2)a_0^2+2(3a_2^2-12a_2+5)a_0+(5a_2-6)a_2+1\left.\right\}\left.\right]=0.\qquad
\end{eqnarray}
Solving for the set of constants \{$a_0$, $a_1$, $a_2$\}, we obtain
\begin{eqnarray}
a_0&=&\frac{C}{2(1-C)}-\chi(\epsilon,C), \nonumber \\
a_1&=&\ln(1-C)-\frac{C}{2(1-C)}+\chi(\epsilon,C), \nonumber \\
a_2&=&-\ln(1-C),\label{eq:const}
\end{eqnarray}
where $\chi$ is known function, but lengthy, fulfills the following limit
\begin{eqnarray}
   \lim_{\epsilon\to 0}\chi(\epsilon,C)=0,
\end{eqnarray}
and the GR limit then is recovered \citep{Roupas:2020mvs}. We note that all the physical quantities within the KB stellar model, i.e. $\bar{\rho}(\epsilon,C)$, $\bar{p}_r(\epsilon,C)$ and $\bar{p}_t(\epsilon,C)$, are given in terms of  two parameters. In this case, observational data provides an extremely important tool to constrain the parameter space \{$\epsilon,C$\} of the matter-geometry nonminimal scenario as suggested by the QRG in comparison with the GR theory. On the other hand, it puts an upper bound on the stellar compactness and consequently its maximum mass.


\section{PSR J0740+6620 Astrophysical Constraints}\label{Sec:constr}


In the following we use NICER measurements of the pulsar PSR J0740+6620 where the hotspot regions are not complicated as in PSR J0030+045 case. Since the pulsar PSR J0740+6620 is in a binary system, its mass can be determined with high accuracy independent from inclination. Using the relativistic Shapiro time delay of ToAs, its mass has been determined as $M= 2.08 \pm 0.07 M_\odot$ \citep{NANOGrav:2019jur,Fonseca:2021wxt}. Since PSR J0740+6620 is a massive pulsar, it is a perfect candidate to examine possible modified gravity scenarios. Recent NICER observations give well constraints on its radius too, it determines mass its radius $R= 13.7_{-1.5}^{+2.6}$ km and $R=12.39_{-0.98}^{+1.30}$ km as indicated by two independent analysis of NICER data \cite{Miller:2021qha,Riley:2021pdl}. It is to be noted that the combined XMM-Newton and NICER dataset enhances the low count rate of the latter, which determines $M=2.07 \pm 0.11 M_\odot$ and $R=12.34^{+1.89}_{-1.67}$ km \citep{Legred:2021hdx}. In our investigation, we use this measurement which is at 68\% CL agreement with \cite{Landry:2020vaw}. In this sense, the pulsar PSR J0740+6620 is a perfect regime to constrain the parameter space of the QRG, i.e. the set of parameters \{$\epsilon$, $a_0$, $a_1$, $a_2$\} or simply \{$\epsilon$, $C$\}.
\subsection{The model parameters}\label{Sec:parameters}
\begin{figure*}
\centering
\subfigure[Mass function]{\label{fig:mass_fn}\includegraphics[scale=0.38]{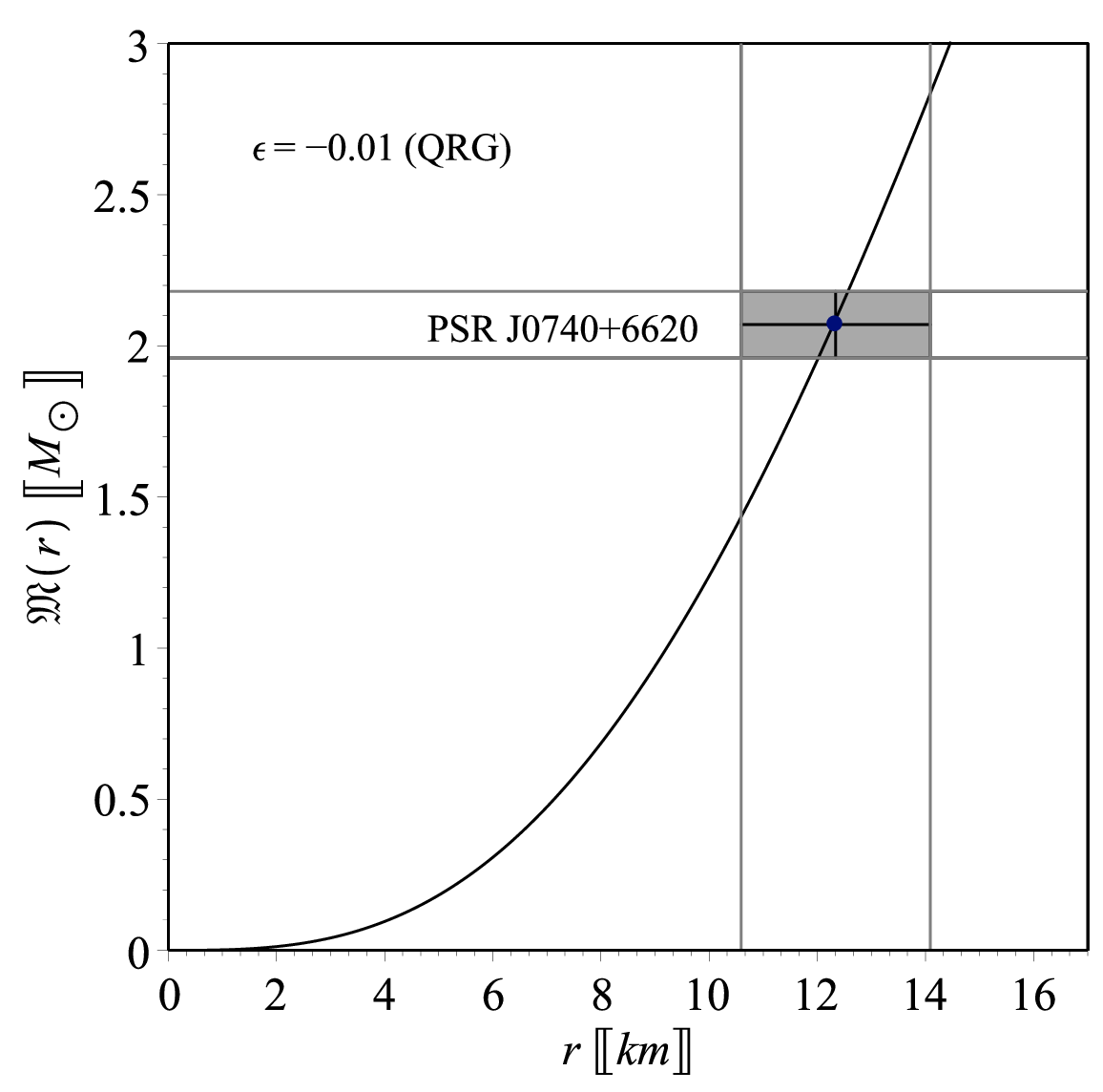}}\hspace{0.5cm}
\subfigure[Matching conditions]{\label{fig:match}\includegraphics[scale=0.38]{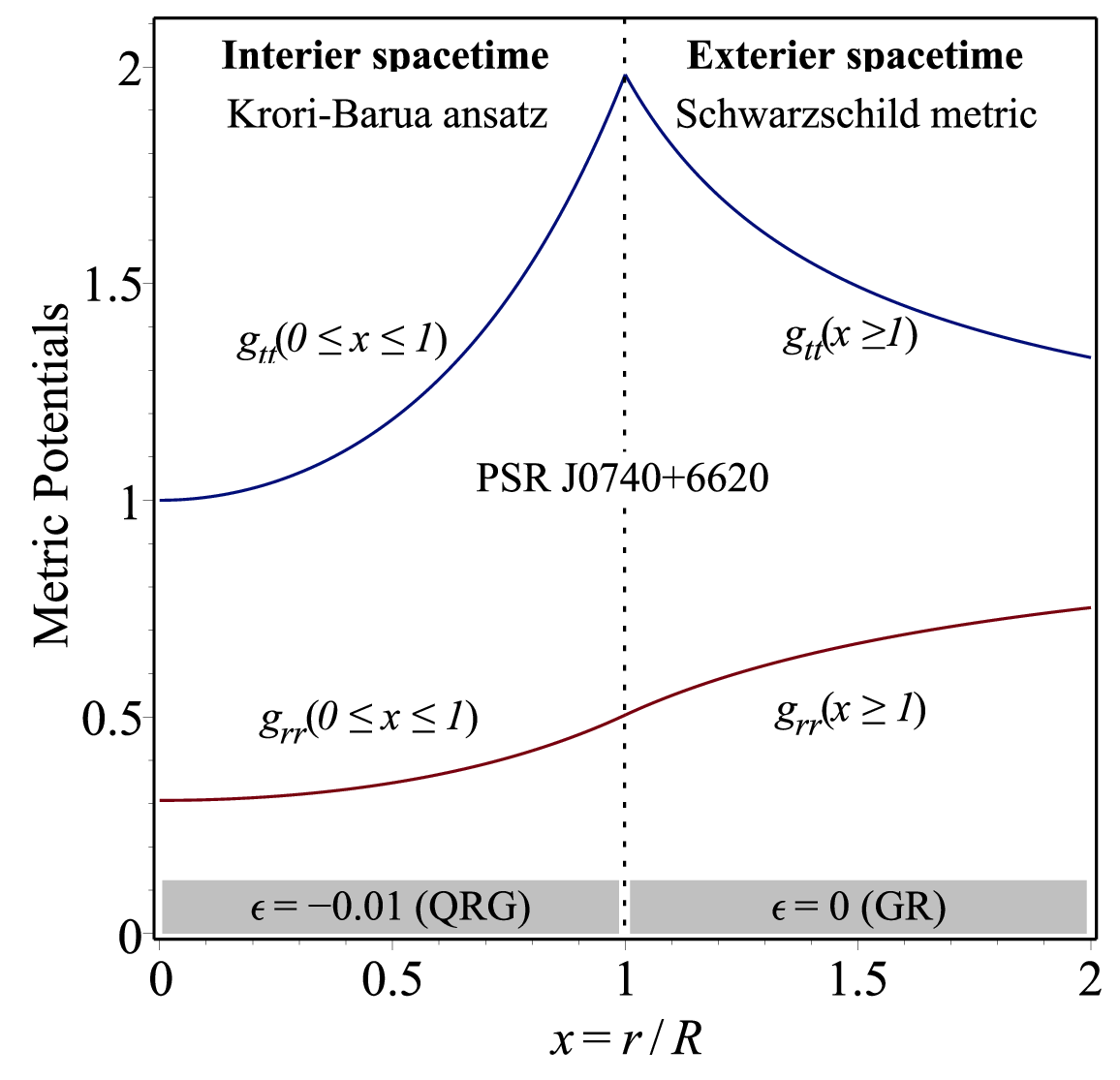}}
\caption{Astrophysical constraints from PSR J0740+6620 observations: \subref{fig:mass_fn} The mass function \eqref{eq:Mass} is in a perfect agreement with NICER+XMM constraints on the mass and radius of PSR J0740+6620, $M=2.07 \pm 0.11 M_\odot$ and $R=12.34^{+1.89}_{-1.67}$ km \citep{Legred:2021hdx}. \subref{fig:match} The metric potentials $g_{tt}$ and $g_{rr}$ consistently match the QRG interior of the pulsar PSR J0740+6620 to the exterior Schwarzschild vacuum where nonminimal coupling vanishes.}
\label{Fig:Matching_conditions}
\end{figure*}

The mean value of the pulsar compactness $C=0.496$. For $\epsilon=-0.01$ and by using the mass function \eqref{eq:Mass}, we estimate the mass of the pulsar PSR J0740+6620 to be $M=2.09 M_\odot$ at radius $R=12.34$ km which is in a perfect agreement with its measured value $M=2.07 \pm 0.11 M_\odot$ and $R=12.34^{+1.89}_{-1.67}$ as seen in Fig. \ref{fig:mass_fn}. Using the matching conditions solution \eqref{eq:const}, we determine the model parameters $a_0 = 0.496$, $a_1 = -1.181$ and $a_2 = 0.684$. In Fig. \ref{fig:match}, we show the regularity of the metric potentials $g_{tt}$ and $g_{rr}$ on the stellar boundary, where KB spacetime describes the stellar interior while the outer region is determined by Schwarzschild vacuum solution. Those numerical values will be used in all figures related to  the pulsar PSR J0740+6620. We note that other possibly choices for $\epsilon>0$ would estimate masses in agreement with observational data. Nevertheless, it would spoil up other physical constraints. This will be explained shortly in Section \ref{Sec:features}.

At the center of PSR J0740+6620 we calculate $\rho(0) = 7.34\times 10^{14}$ [g/cm$^3$] $\approx 2.72 \rho_\text{nuc}$ and $p_r(0) = p_t(0)= 8.72\times 10^{34}$ [dyn/cm$^2$].
At the surface $\rho(R) = 4.19\times 10^{14}$ [g/cm$^3$] $\approx 1.55 \rho_\text{nuc}$, $p_r(R) = 0$ [dyn/cm$^2$] and $p_t(R) = 3.27\times 10^{34}$ [dyn/cm$^2$]. Also we calculate the redshift $Z(r):=-1+1/\sqrt{-g_{tt}}$ at the pulsar center $Z(0)=0.80$ and at its surface $Z(R)=0.41$. In comparison to GR we obtain $\rho(0) = 7.22\times 10^{14}$ [g/cm$^3$] $\approx 2.68 \rho_\text{nuc}$, $p_r(0) = p_t(0)= 9.43\times 10^{34}$ [dyn/cm$^2$] and $Z(0)=0.80$. While at the surface we obtain $\rho(R) = 4.17\times 10^{14}$ [g/cm$^3$] $\approx 1.55 \rho_\text{nuc}$, $p_r(R) = 0$ [dyn/cm$^2$] and $p_t(R) = 3.24\times 10^{34}$ [dyn/cm$^2$] and $Z(R)=0.41$. Although both QRG and GR predict almost same densities, pressures and redshifts some other physical properties of high importance are significantly different. The next subsection is devoted to discuss this issue.
\subsection{Speed of sound and conformal anomaly}\label{Sec:conformal}
One of the basic quantities which determines the interior of the compact star is the adiabatic sound speed. Particularly, in the radial direction
\begin{equation}\label{eq:sound_speed}
  v_r^2 = \frac{dp_r}{d\rho}=\frac{\bar{p}'_r}{\bar{\rho}'}.
\end{equation}
Obviously, it is constrained by causality and thermodynamic stability as $0\leq v_r^2 \leq 1$. On the other hand, it is strongly related to the EoS as discussed in Subsec. \ref{Sec:EoS}. It is well known that the sound speed square $v_r^2 \ll c^2$ of non-relativistic matter at densities much lower than $\rho_\text{nuc}$, while it approaches the conformal upper bound $v_r^2=c^2/3$ at much high densities (ultra-relativistic matter) $\rho \sim 40 \rho_\text{nuc}$ where perturbative quantum chromodynamics (pQCD) approach is applied. At densities few times $\rho_\text{nuc}$ which is compatible with compact star density, two scenarios are suggested: The conjectured conformal upper limit is broken where the sound speed increases non-monotonically to a peak and decreases toward the conformal sound speed limit from below at high densities. The other scenario is to keep the conformal constraint at all densities where the sound speed increases monotonically to approach the upper limit $v_r^2/c^2\to 1/3$. The first scenario is suggested to explain existence of massive pulsars, $M \gtrsim 2 M_\odot$, where the sound speed peaks up to $v_r^2 \sim 0.7 c^2$ at densities  $\rho \sim 3-5 \rho_\text{nuc}$ somewhere inside the star (for an isotropic perfect fluid). However, the tidal deformability of the observed gravitational wave signals of NS-NS merging GW170817 and GW190425 sets an upper bound on the stiffness of the EoS inconsistent with the first scenario.

\begin{figure*}[t]
\centering
\subfigure[Sound speed]{\label{fig:sound}\includegraphics[scale=0.38]{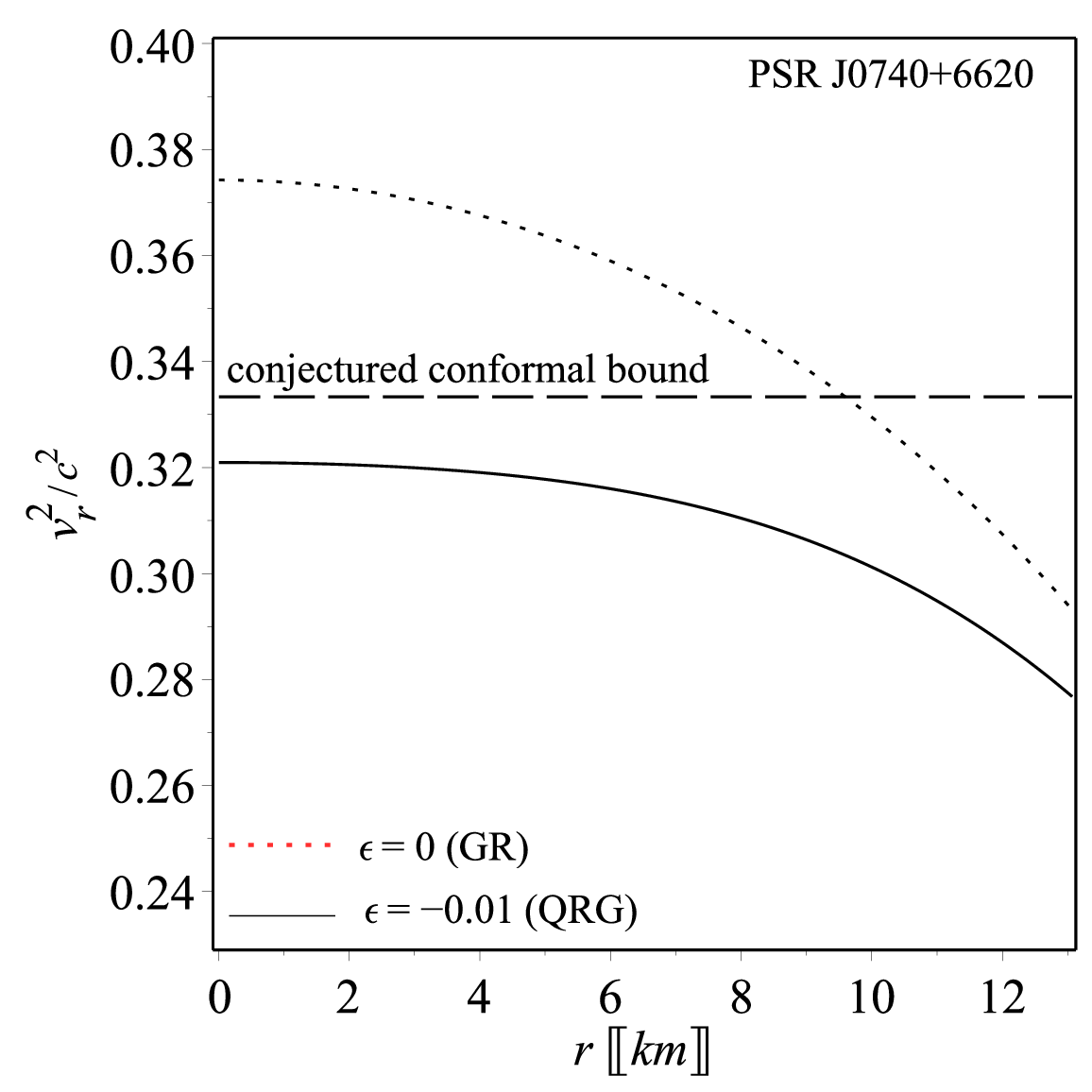}}\hspace{0.5cm}
\subfigure[Conformal anomaly]{\label{fig:trace_a}\includegraphics[scale=0.38]{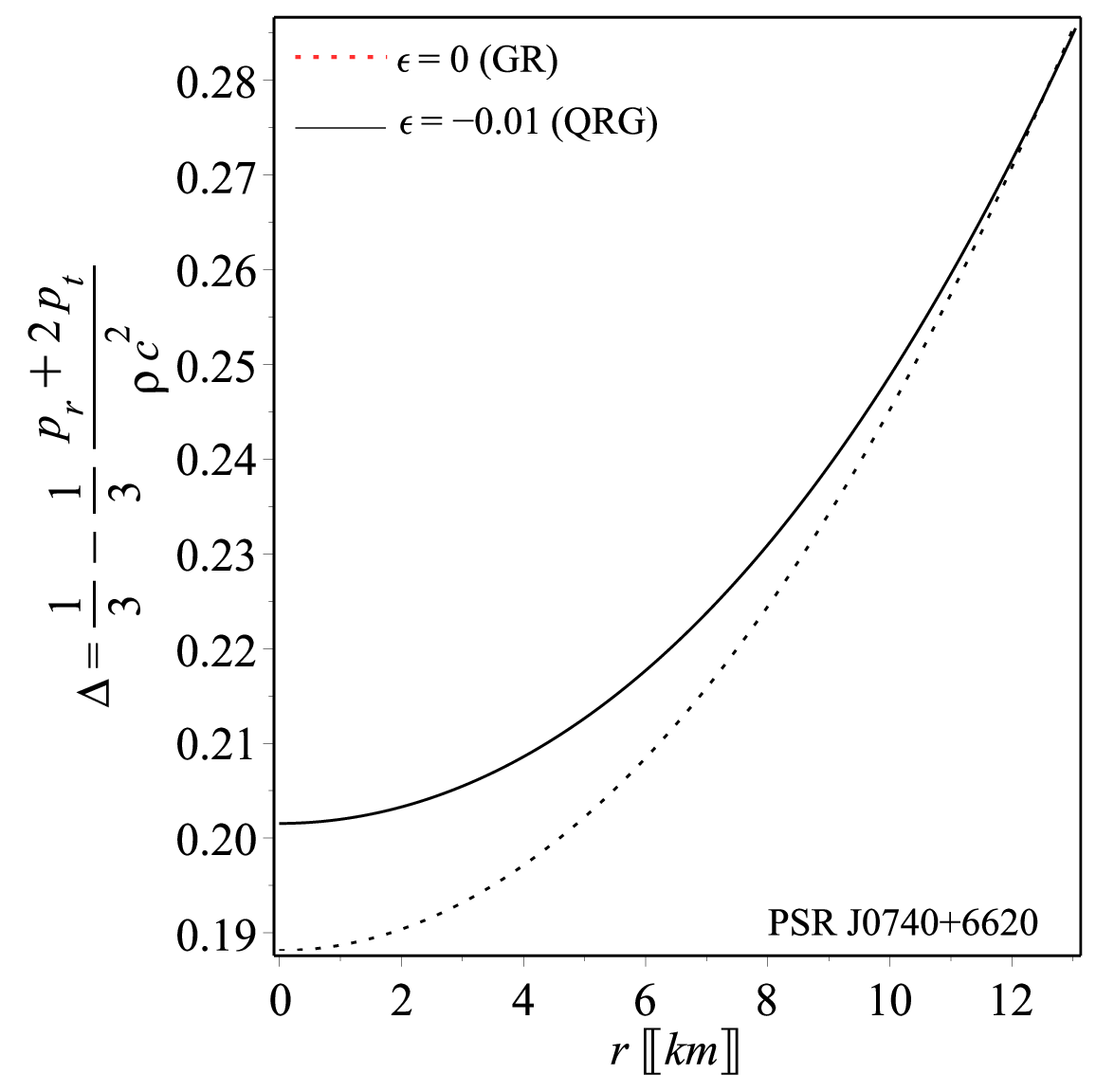}}
\caption{Distinguished features of QRG: \subref{fig:sound}  The radial dependence of the adiabatic sound speed \eqref{eq:sound_speed} inside the pulsar  PSR J0740+6620. The conformal upper limit of the sound speed, $v_r^2/c^2\leq 1/3$, holds everywhere when nonminimal coupling does not vanish. This distinguishes the QRG from GR in which the conformal limit is violated for radial distance $0\leq r \lesssim 9.59$.  \subref{fig:trace_a} The radial dependence of the trace anomaly \eqref{eq:conf_anom} shows that $\Delta>0$ everywhere which indicates that the conformal symmetry is broken inside the pulsar PSR J0740+6620. It decreases monotonically toward the center where $\Delta$ gets higher values than in GR systematically toward the stellar center.}
\label{Fig:conformal}
\end{figure*}

In fact, we show that the conjectured conformal sound speed can applied everywhere, according to the present model, inside such a heavy object as in the case of the compact star PSR J0740+6620. Using \eqref{eq:Feqs3} and \eqref{eq:sound_speed}, we plot the radial dependence of adiabatic sound speed within the pulsar PSR J0740+6620 as shown in Fig. \ref{Fig:conformal}\subref{fig:sound}.  Unlike the case when GR is applied, the QRG predicts a sound speed consistent with the conjecture conformal limit everywhere inside the pulsar where $v_r^2/c^2\sim 0.28$ at the star surface and $v_r^2/c^2\sim 0.32$ at its inner core. Remarkably when the nonminimal coupling scenario is dropped ($\epsilon=0$) the conformal sound speed is mostly no longer hold for a radial distance $0\leq r \lesssim 9.59$ km, where the sound speed approaches $v_r^2/c^2 \simeq 0.37$ at the stellar center. We interpret the obtained result as follows: The strong anisotropic conditions ($p_t>p_r$) induces a repulsive force to oppose the gravitational collapse and in turn allows for more mass with no need for higher stiffness (large sound speed). In fact, this force cannot reduce the sound speed enough as already obtained for the GR case. The nonminimal coupling, however, contributes as an additional regulating repulsive force holding the sound speed below the conformal limit. This argument will be discussed in more detail in the following subsection.

Another physical quantity which is related to our approach is the conformal trace anomaly
\begin{equation}\label{eq:conf_anom}
  \Delta:=\frac{1}{3}\frac{g_{\alpha\beta} \mathfrak{T}^{\alpha\beta}}{\rho c^2}=\frac{1}{3}\left[1-\frac{p_r+2p_t}{\rho c^2}\right],
\end{equation}
which provides a dimensionless measure of possible deviation from conformal symmetry. At high densities, pQCD exhibits conformal symmetry where $\Delta\to 0$. At finite densities and/or temperatures, which is compatible with NS densities, it has been argued that the conformal symmetry is broken due to running of the strong coupling with energy at quantum level where $\Delta \neq 0$ \cite{Fujimoto:2022ohj}. In general, thermodynamic stability and causality put a constraint on the trace anomaly $-2/3 \leq \Delta \leq 1/3$, whereas its value and sign inside NSs are not known. We remind that the matter-geometry nonminimal coupling in QRG is motivated by the trace anomaly when the backreaction of the quantum fields to a curved spacetime geometry is considered as suggested by \eqref{eq:QRT}. Using \eqref{eq:Feqs3} and \eqref{eq:conf_anom}, we plot the radial dependence of trace anomaly within the pulsar PSR J0740+6620 as shown in Fig. \ref{Fig:conformal}\subref{fig:trace_a}. Clearly, the conformal trace anomaly $\Delta$ is positive everywhere inside the pulsar, which means that the conformal symmetry is broken inside the pulsar. It approaches the vacuum value $\Delta \to 1/3$ at the pulsar surface and monotonically decreases toward $\Delta \to 0.2$ at the center of the pulsar. Similar behaviour has been obtained for typical NS with mass $M\sim 1.4 M_\odot$ \cite{Ecker:2022dlg}. We note that the trace anomaly at the surface is not exactly $\Delta(R)=1/3$, since $\rho(R)\neq 0$ and $p_t(R)\neq 0$. Therefore, it may point out that the pulsar PSR J0740+6620 is self-bound. Similar behaviour is shown when the nonminimal coupling vanishes ($\epsilon=0$), it also shows that the QRG exhibits more positivity of the trace anomaly than GR. This will affect the maximum compactness allowed by both theories.
\subsection{Hydrostatic equilibrium and thermodynamic stability}\label{Sec:equilibrium}
\begin{figure*}[t]
\centering
\subfigure[Hydrostatic Equilibrium]{\label{fig:forces}\includegraphics[scale=0.38]{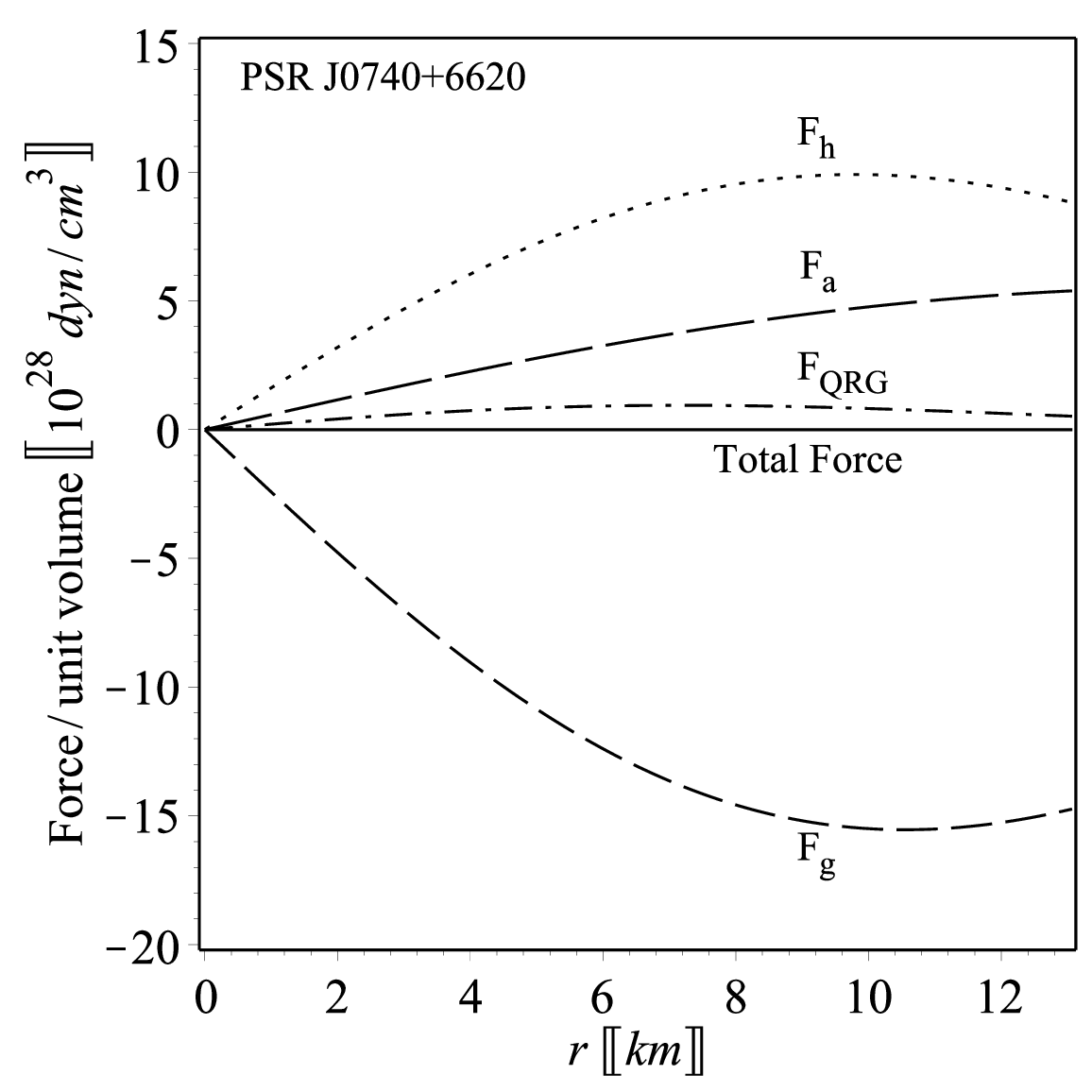}}\hspace{0.5cm}
\subfigure[Thermodynamic Stability]{\label{fig:adiabatic}\includegraphics[scale=0.38]{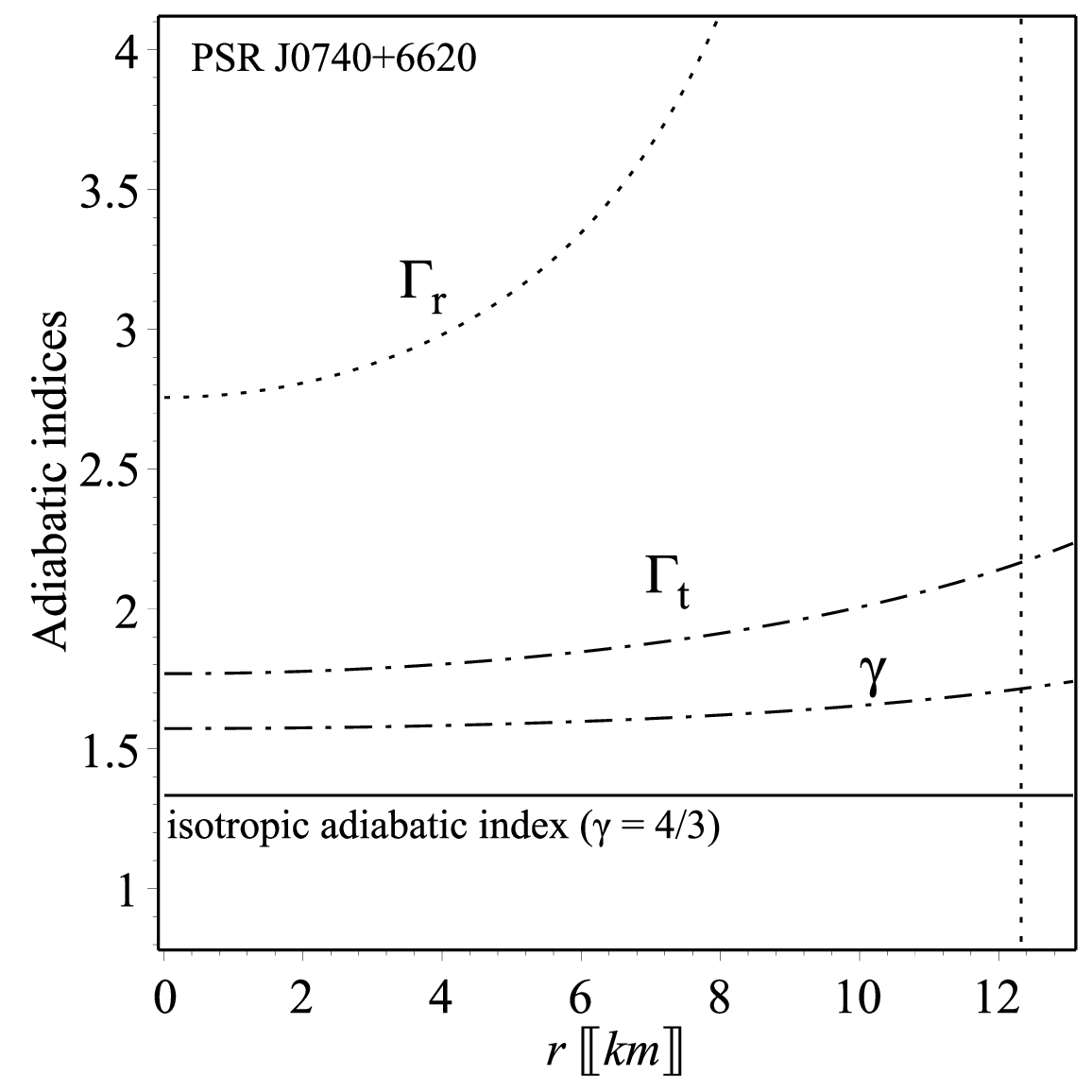}}
\caption{Hydrostatic equilibrium and thermodynamic stability of the pulsar PSR J0740+6620: \subref{fig:forces} The contributing forces in modified TOV equation of hydrostatic equilibrium \eqref{eq:TOV}. The newly introduced force due to nonminimal coupling, $F_\text{QRG}$, is clearly repulsive which acts in addition to anisotropic  and hydrostatic forces, $F_a$ and $F_h$, to balance the collapsing gravitational force, $F_g$. \subref{fig:adiabatic} The radial dependence of the adiabatic indices \eqref{eq:adiabatic} verifies that the pulsar is stable since $\gamma>4/3$ and $\Gamma_r,~\Gamma_t> \gamma$. The divergence of the adiabatic index $\Gamma_r$ at the pulsar surface characterizes self-bound objects which allows for very stable massive compact stars even with low sound speed.}
\label{Fig:equilbrium}
\end{figure*}

The assumption $\nabla_{\alpha}\widetilde{\mathfrak{T}}{^\alpha}{_\beta}=0$ constrains the spherically symmetric star of anisotropic fluid to be in static gravitational equilibrium according to the quadratic nonminimal coupling model. This modifies the Tolman–Oppenheimer–Volkoff (TOV) equation to read
\begin{equation}\label{eq:TOV}
    p'_r=-\frac{1}{2}\alpha'(\rho c^2+p_r)+2\frac{\delta}{r}+\widetilde{B}'.
\end{equation}
where $\widetilde{B}=2{\epsilon  \kappa \ell^2} \left[\tilde{\rho} c^2 (\tilde{p}_r+2\tilde{p}_t)-\tilde{p}_t(\tilde{p}_t+2\tilde{p}_r)\right]$. The left hand side of the above equation represents a repulsive force--given that the radial pressure gradient is negative everywhere inside the star-- due to hydrostatic pressure $p'_r=:-F_h$. The first term of the right hand side represents the relativistic gravitational force $F_g:=-\frac{1}{2}\alpha'(\rho c^2+p_r)=-\frac{a_0 r}{R^2}(\rho c^2+p_r)$, which is attractive in its nature as the density and the pressure are positive everywhere inside the star, and clearly $a_0$ must be positive. The second and the third terms, $F_a:=2\delta/r=2(p_t-p_r)/r$ and $F_{QRG}:=\widetilde{B}'$, represent two additional forces due the anisotropy and the quadratic nonminimal coupling contributions to the hydrostatic equilibrium. Therefore, we write the balanced forces $F_h+F_g+F_a+F_{QRG}=0$ which are represented by Fig \ref{Fig:equilbrium}\subref{fig:forces} for the interior of the pulsar PSR J0740+6620.

The figure clearly shows that the collapsing (attractive) gravitational force is balanced by the repulsive forces where the anisotropic (repulsive) force $F_a>0$ due to strong anisotropic condition $p_t>p_r$, similarly, for the matter-geometry nonminimal coupling, it introduces another repulsive force $F_{QRG}$ which plays a key role to allow for higher values of the maximum compactness in QRG than GR (see Sec. \ref{Sec:max_comp}) and also to determine the size of the star (see Sec. \ref{Sec:J0952–0607}).

Another physical condition on the neutron star is to examine its stability against adiabatic perturbation. This can be provided via some physical constraints on the relativistic adiabatic indices
\begin{equation}\label{eq:adiabatic}
\gamma=\frac{4}{3}\left(1+\frac{F_a}{2 |p'_r|}\right)_{max},\,
\Gamma_r=\frac{\rho c^2+p_r}{p_r} (v_r/c)^2,\,
\Gamma_t=\frac{\rho c^2+p_t}{p_t} (v_t/c)^2.
\end{equation}
For isotropic spheres, i.e. the anisotropy function $\delta=0$, we obtain $\gamma=4/3$ and $\Gamma_t=\Gamma_r$. In the general case of anisotropic fluid, the star is in neutral equilibrium where $\Gamma=\gamma$, while it is in a stable equilibrium where $\Gamma>\gamma$ everywhere in the star interior.

Using \eqref{eq:Feqs3} and \eqref{eq:adiabatic}, we plot the radial dependence of adiabatic indices within the pulsar PSR J0740+6620 as shown in Fig. \ref{Fig:equilbrium}\subref{fig:adiabatic}. It can be figured out that $\gamma>4/3$ inside the star which characterizes strong anisotropy cases with $p_t > p_r$, in addition $\Gamma_r > \gamma$ and $\Gamma_t > \gamma$ which insure that the pulsar is in stable equilibrium against radial perturbation \cite{1975A&A....38...51H,10.1093/mnras/265.3.533}. Remarkably, as shown in the figure, the divergence of the radial adiabatic index at the surface of the pulsar PSR J0740+6620, in general, characterizes self-bound objects and allows for very stable heavy stars $\simeq 2 M_\odot$ even if the sound speed is small \cite{Annala:2019puf}. This could provide an evidence for quark matter in massive stars such as the PSR J0740+6620. Although the present approach is completely geometric, we showed that the KB ansatz \eqref{eq:KB} can be used to relate radial/tangential pressures to density in linear relations as given by \eqref{eq:EoSx}. In the following subsection we investigate the EoS which describes the interior of the pulsar PSR J0740+6620.
\subsection{KB induced EoS}\label{Sec:KBEoS}
\begin{figure*}
\centering
\subfigure[Radial EoS]{\label{fig:REoS}\includegraphics[scale=0.4]{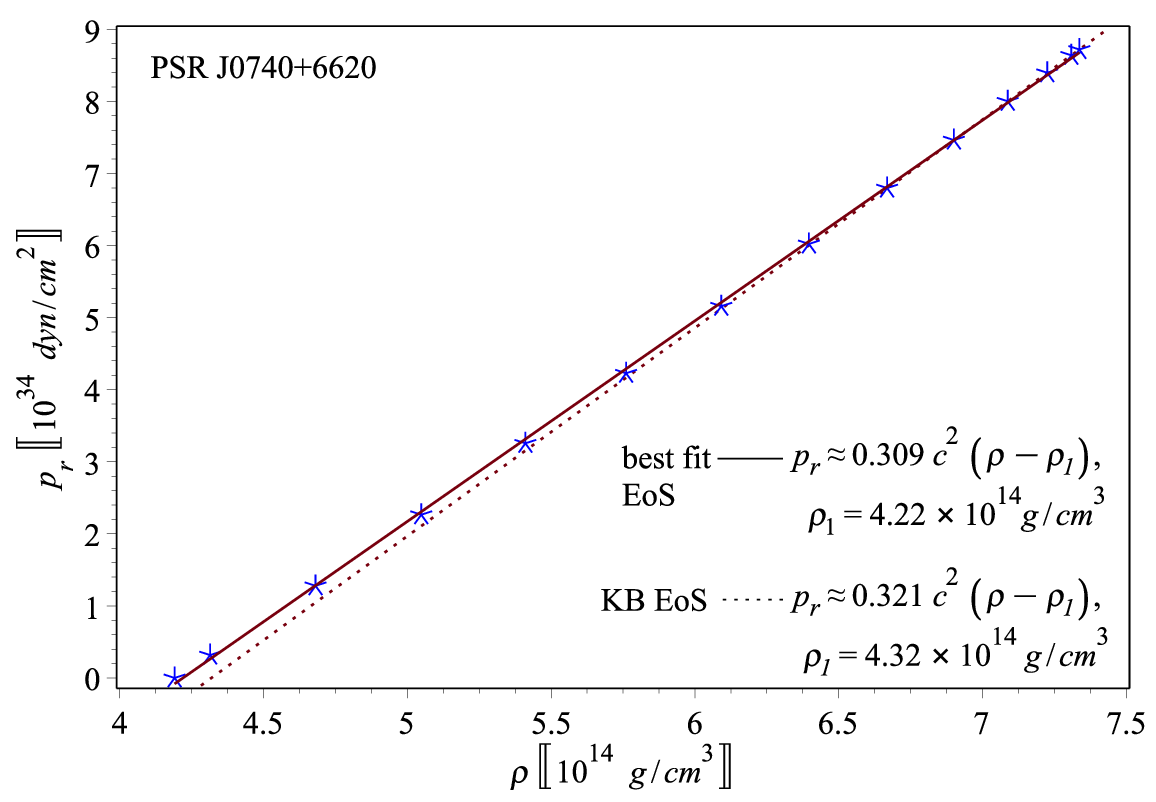}}\hspace{0.1cm}
\subfigure[Tangential EoS]{\label{fig:TEoS}\includegraphics[scale=0.4]{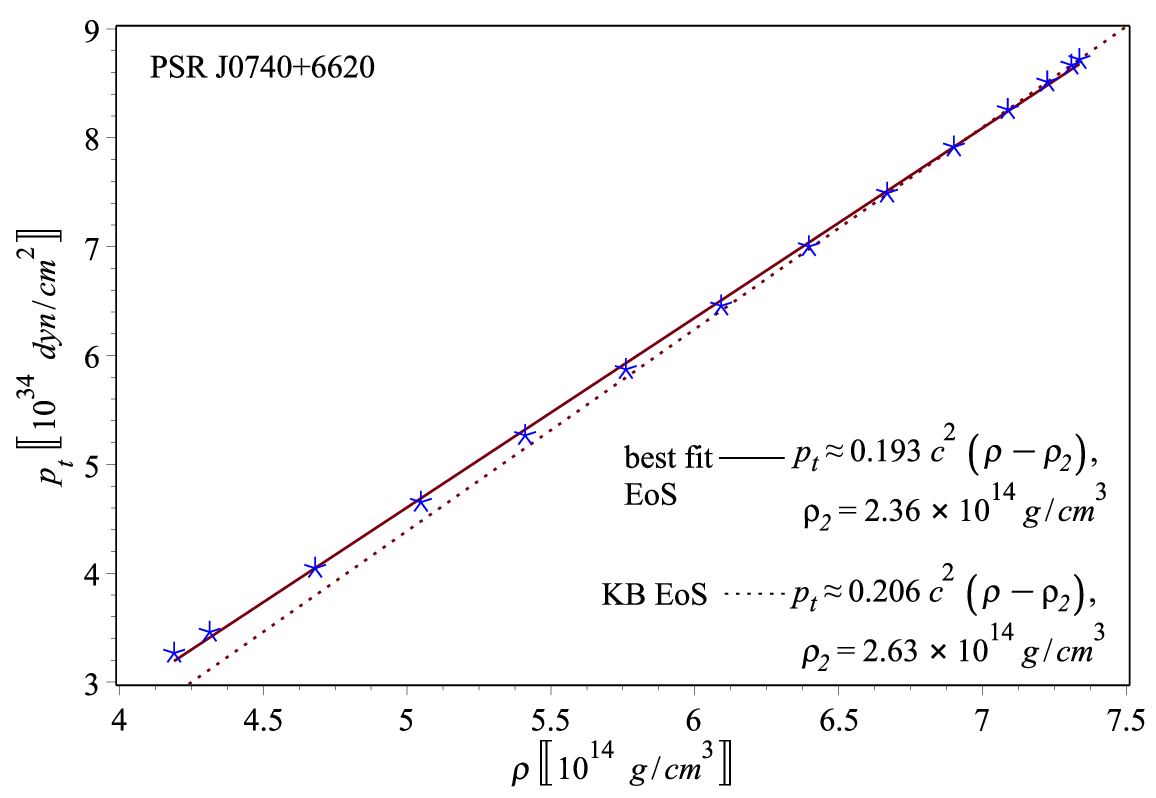}}
\caption{EoSs of the pulsar PSR J0740+6620: A comparison between the KB EoSs\eqref{eq:J0740KBEoS}  and the best-fit EoSs
\eqref{eq:J0740bestEoS}. Both coincide at the stellar center while slight systematic deviations appear toward the surface, which should be understood due to KB series approximation \eqref{eq:EoSx}. We plot the radial EoSs $p_r(\rho)$ in Fig. \subref{fig:REoS} and the tangential EoSs $p_t(\rho)$ in Fig. \subref{fig:TEoS}.}
\label{Fig:EoS}
\end{figure*}

The EoS of the matter inside compact stars is the ultimate goal of  several studies. For this reason we rewrite \eqref{eq:EoSx} in more physical form
\begin{equation}\label{eq:EoS}
    p_r(\rho)\approx v_r^2(\rho -\rho_1), \quad p_t(\rho)\approx v_t^2(\rho - \rho_2),
\end{equation}
where $c_1=v_r^2/c^2$ and $c_0=-c_1(\rho_1/\rho_\star)$, and similarly  $\tilde{c}_1=v_t^2/c^2$ and $\tilde{c}_0=-\tilde{c}_1(\rho_2/\rho_\star)$. Since equations \eqref{eq:EoS} are induced by KB ansatz \eqref{eq:KB}, we refer to them as KB EoSs. For PSR J0740+6620 case, we evaluate $c_0=-0.395$, $c_1=0.321$, $\tilde{c}_0=-0.154$, $\tilde{c}_1=0.206$ and $\rho_\star=3.52\times10^{14}$ g/cm$^3$. This directly gives $v_r^2=0.321 c^2$, surface density $\rho_s=\rho_1=4.32\times 10^{14}$ g/cm$^3$, $v_t^2=0.206 c^2$ and $\rho_2=2.63\times 10^{14}$ g/cm$^3$, which in turn provides the following EoSs
\begin{eqnarray}\label{eq:J0740KBEoS}
    p_r&=&0.32 c^2 (\rho-4.32\times 10^{14}) [\text{dyn/cm}^2],\nonumber\\[5pt]
    p_t&=&0.21 c^2 (\rho-2.63\times 10^{14}) [\text{dyn/cm}^2].
\end{eqnarray}
We note that the above EoSs are mostly valid at the center of the pulsar as required by the series approximation as discussed in Subsec. \ref{Sec:EoS}. Therefore, the sound speed is identical to the exact value as obtained in Subsec. \ref{Sec:conformal}, whereas the surface density according to the above calculations is slightly higher than the exact value $\rho(R) = 4.19\times 10^{14}$ [g/cm$^3$] $\approx 1.55 \rho_\text{nuc}$ as obtained in Subsec. \ref{Sec:parameters}. On the other hand, we generate a sequence of exact values of the pressures and density from the center to the surface of the pulsar PSR J0740+6620, as labeled by blue asterisk on the plots of Fig. \ref{Fig:EoS}, then we give the best fit EoSs
\begin{eqnarray}\label{eq:J0740bestEoS}
    p_r&=&0.31 c^2 (\rho-4.22\times 10^{14}) [\text{dyn/cm}^2],\nonumber\\[5pt]
    p_t&=&0.19 c^2 (\rho-2.36\times 10^{14}) [\text{dyn/cm}^2].
\end{eqnarray}
Those are in a perfect agreement with equations \eqref{eq:J0740KBEoS}, which proves the validity of KB EoSs not only at the center of the pulsar but also at every point inside it. We plot both KB and the best fit EoSs as shown by Fig. \ref{Fig:EoS} to verify our numerical calculations. Both coincide at the center with slight deviations toward the surface of the pulsar as expected.

We summarize our findings as follows: The observational constraints on the mass and radius of the pulsar PSR J0740+6620 provided a perfect laboratory to estimate the amount of the matter-geometry nonminimal according to the QRG approach. It determines the dimensionless nonminimal coupling parameter $\epsilon=-0.01$. Accordingly, we determined the contribution of the additional repulsive force due to QRG to the hydrostatic equilibrium. This allows to reduce the maximum sound speed at the center to fulfil its conformal upper limit, $v_r^2/c^2=1/3$, which significantly differentiates the QRG from GR. Moreover, we showed that the conformal symmetry is broken everywhere where the trace anomaly $\Delta$ is always positive, in particular $0.20 \lesssim \Delta \lesssim 0.28$ for the pulsar PSR J0740+6620. Furthermore, we showed that the pulsar is stable against radial pulsation since the adiabatic index $\Gamma_r>\gamma>4/3$. In conclusion, the divergence of the radial adiabatic index at the surface and the validity of the conformal sound speed suggest that the pulsar is self-bound where the stellar interior consists of quark matter. More physical features of the QRG and its relevance to strange quark matter is investigated in the following section.


\section{Physical Features}\label{Sec:features}

In the present section, we justify our choice of the negative sign of the dimensionless parameter $\epsilon$ which determines the amount of matter-geometry nonminimal coupling due to QRG. In addition, we investigate the maximum compactness allowed by QRG in comparison to the corresponding value in GR theory. Also, we investigate possible role of matter-geometry nonminimal coupling relevant to the microscopic structure of the star. We note that the results included in the present section are valid for QRG with $\epsilon=-0.01$, in general, independent from observational data.
\subsection{negative value of the nonminimal coupling parameter}\label{Sec:neg_para}
Our choice of the sign of the parameter $\epsilon$ is not yet justified, since positive values can produce masses compatible with NICER+XMM-Newton constraints on the pulsar PSR J0740+6620 too. However, one of the direct consequences of non negative $\epsilon$ is the violation of conformal constraints on the sound speed upper limit. For example, for $\epsilon=+0.01$, we obtain $v_r^2\simeq 0.43 c^2$ at the stellar center even higher than it corresponding GR value $v_r^2\simeq 0.37 c^2$ ($\epsilon=0$) as shown in Fig. \ref{fig:sound}. We interpret this result as the nonminimal coupling force in the hydrostatic TOV equation becomes negative (collapsing) force for $\epsilon>0$. In turn, it would squeeze the pulsar mass to smaller sizes with stiffer EoS. Other consequences on the maximum allowed compactness will be mentioned shortly in the next subsection.

In order to verify our conclusion, we perform the following simple calculations. If the conformal sound speed is hold within PSR J0740+6620 according to our claim, and by knowing only its measured mass $M=2.07\pm 0.11 M_\odot$, the estimated radius then varies according to the sign of $\epsilon$ as provided by Table \ref{Tab:neg_para}. In the last column, we calculate the tension level between the estimated radius, for different $\epsilon$ values, and the measured one, $R=12.34\pm1.75$ km, in terms of the standard deviation $\sigma$. Clearly, the conjectured conformal sound speed excludes $\epsilon\geq 0$ up to $\gtrsim 1.6\sigma$, whereas positive $\epsilon$ is in a stronger tension, $\sim 2.85\sigma$, with observational data.
\begin{table}[h!]
    \caption{Estimated radius, $R_{est}$, of the pulsar PSR J0740+6620 with $M=2.07\pm 0.11 M_\odot$ for different $\epsilon$-values assuming the conformal constraint on the sound speed is hold, by taking $v_r^2/c^2=0.32$ at the stellar core. The KB model parameters \{$a_0$, $a_1$, $a_2$\} are listed for each case. The last column shows the tension level, in terms of the standard deviation $\sigma$, with the observed radius $R_{obs}= 12.34\pm1.75$ km \citep{Legred:2021hdx}.}
    \label{Tab:neg_para}
    \centering
    \begin{ruledtabular}
        \begin{tabular}{cccccc}
            $\epsilon$    & $a_0$  & $a_1$ & $a_2$ & $R_{est}$ [km]& Tension [$\sigma$]  \\
        \hline
        $-0.01$  & $0.49$ &  $-1.17$   & $0.68$ & $12.44 \pm 0.66$ & $0.05$\\
        $0$      & $0.33$ &  $-0.83$   & $0.50$ & $15.46 \pm 0.82$ & $1.61$\\
        $+0.01$  & $0.26$ &  $-0.67$   & $0.41$ & $18.03 \pm 0.96$ & $2.85$
            \end{tabular}
     \end{ruledtabular}
\end{table}

As we mentioned earlier in Sec. \ref{Sec:QRgrav}, our choice of the matter-geometry nonminimal term in QRG is motivated by the trace anomaly where backreaction of the quantum fields to a curved spacetime geometry is considered. We note that the quantum corrections due to trace anomaly of Type A and B are in general proportional to quadratic curvature invariants.  In particular, the coefficient of Euler density contribution--similar to QRG-- to the trace anomaly is explicitly \citep{Duff:1993wm}
\begin{equation}
    \beta= - \frac{1}{360(4\pi)^2}\left(N_s+11 N_f+62 N_\nu\right),
\end{equation}
where $N_s$, $N_f$ and $N_\nu$ are the number of scalars, Dirac fermions and vector fields. As noted in \cite{Awad:2015syb}, the coefficient $\beta$, in general, must be negative. The results of Table \ref{Tab:neg_para} along with the above comments on the trace anomaly quantum corrections may indicate an underlying similarity between QRG approach and trace anomaly by realizing the mutual role of $\beta$ and $\epsilon$ (precisely $\epsilon \ell^2$), if any. It should be understood, however, the trace anomaly needs a completely different setup which may manifest solutions with different characteristics. We are currently investigating this issue.

\subsection{Trace energy condition and maximum compactness}\label{Sec:max_comp}

We have shown that the conformal symmetry is broken inside the pulsar PSR J0740+6620. This has been featured via nonvanishing trace anomaly \eqref{eq:conf_anom}. Indeed the positivity of the trace anomaly $\Delta$ as obtained in the present study indicates the trace of the stress-energy tensor is negative everywhere inside the pulsar, i.e. $\mathfrak{T}=p_r+2p_t-\rho c^2 <0$, which we refer to as the trace energy condition (TEC). It has been argued that this inequality could be violated in the most massive NS cores \citep{Podkowka:2018gib}. Although this is not widely accepted, it could be true when the sound speed violates the conformal upper bound, i.e.  $v_r^2/c^2>1/3$, see the given example by Zeldovich which violates the TEC while $v_r\simeq c$ \cite{osti_4803023}. This may indicate that the medium is not at its true ground state similar to the case of interference of sound waves when high-frequencies are amplified and low-frequencies are damped. In the present study, we verified that $v_r^2/c^2 \lesssim 1/3$ and $\Delta \geq 0$, consequently the TEC must hold everywhere inside the pulsars.
\begin{figure*}
\centering
\subfigure[Trace energy condition]{\label{fig:TEC1}\includegraphics[scale=0.4]{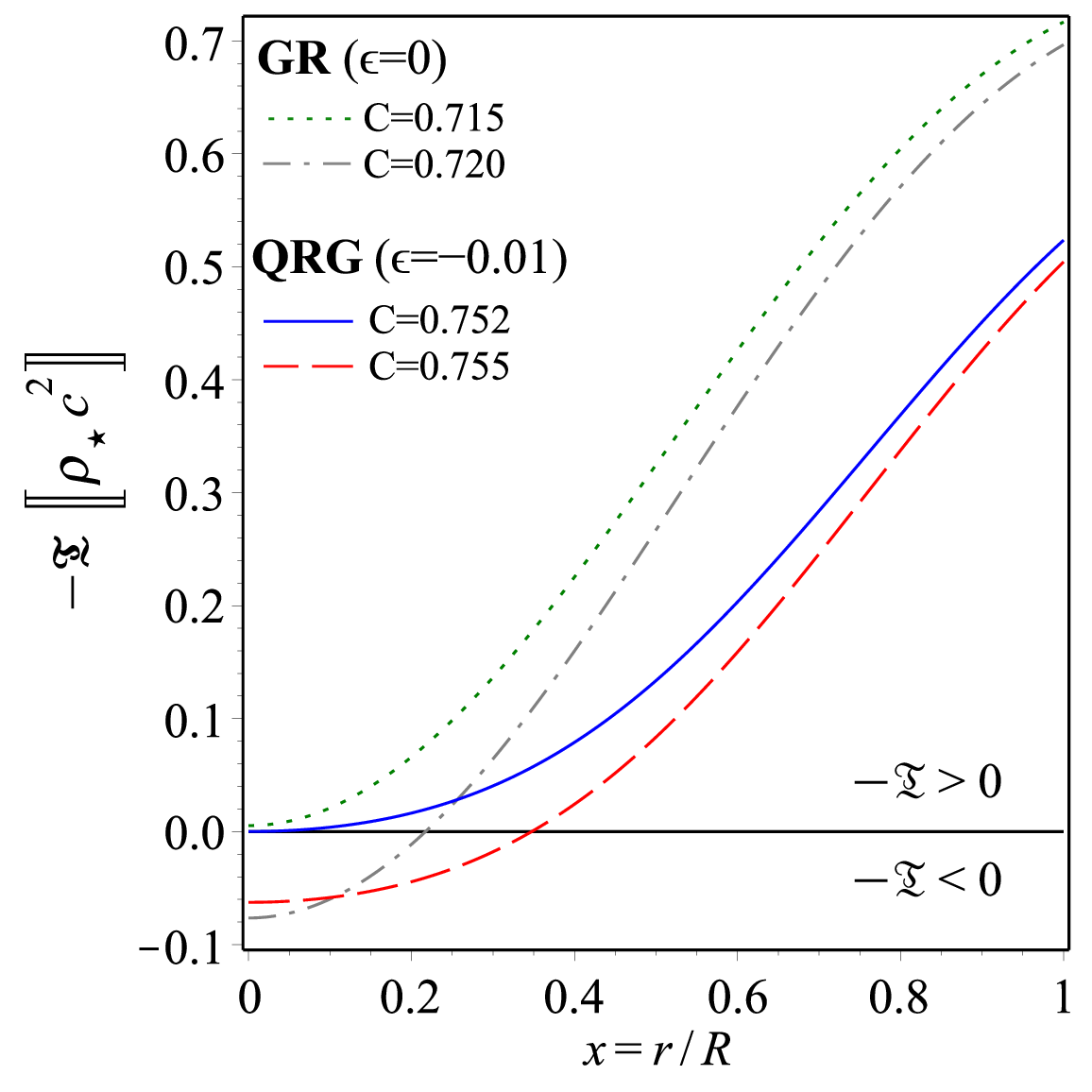}}\hspace{0.3cm}
\subfigure[Trace-Compactness diagram]{\label{fig:TEC2}\includegraphics[scale=0.4]{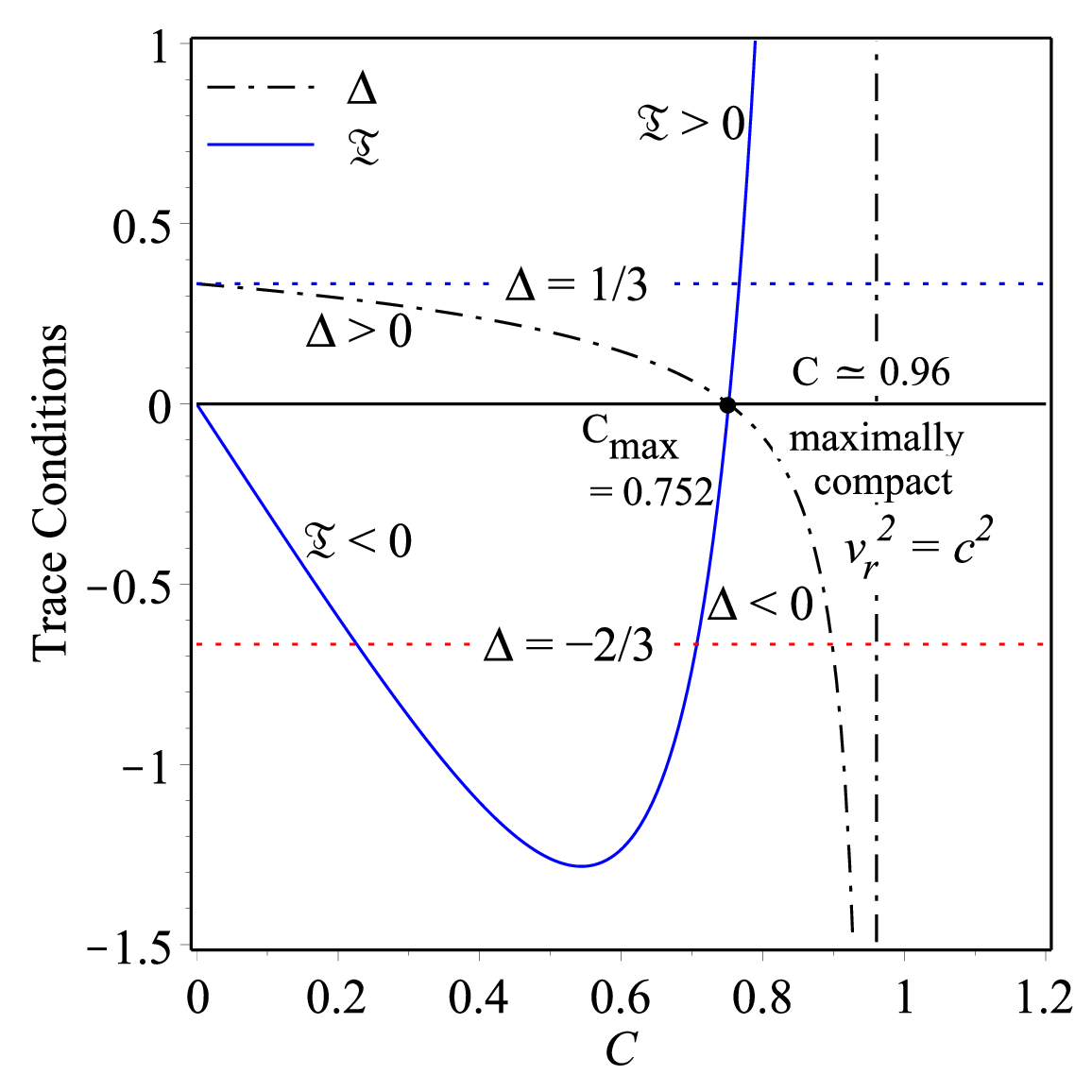}}
\caption{Maximum compactness: \subref{fig:TEC1} Using the field equations and the KB parameters \eqref{eq:Feqs3} and \eqref{eq:const} with $\epsilon=-0.01$, we plot the radial dependence of the trace $\mathfrak{T}=p_r+2p_t-\rho c^2$ in terms of the characteristic energy-density $\rho_\star c^2=\kappa^{-1} R^{-2}$. The TEC, $\mathfrak{T}\leq 0$, is verified everywhere inside the star, whereas the maximum compactness accordingly is $C_\text{max}\simeq 0.752$ where $\mathfrak{T}=0$ at the stellar center, which is $\sim 4\%$ higher than the corresponding value in GR ($\epsilon=0$). \subref{fig:TEC2} We combine two phase-spaces, namely $\Delta-C$ and $\mathfrak{T}-C$, by applying QRG with $\epsilon=-0.01$ at the stellar center, which determine the maximum compactness $\simeq 0.752$ according to the constraints $\Delta=\mathfrak{T}=0$. The $\Delta-C$ diagram shows that $\Delta\to -\infty$ as $C\to 0.96$. Same value of $C$ can be obtained for a maximally compact object with EoS, $p_r=v_r^2(\rho-\rho_s)$, where $v_r=c$.}
\label{Fig:TEC}
\end{figure*}

We utilize the TEC to set a restrictive constraint on the maximum compactness allowed by the QRG scenario. Using the field equations \eqref{eq:Feqs3} where the KB parameters are given by \eqref{eq:const}, we obtain that the maximum compactness corresponds to QRG with $\epsilon=-0.01$ is $C_\text{max}=0.75$ which is $\sim 4\%$ higher than the GR case \citep{Roupas:2020mvs}. This results is expected as we pointed out in Subsec. \ref{Sec:conformal}, where the trace anomaly $\Delta$ is almost the same in QRG and GR at the surface, while in QRG it systematically gets higher positive values toward the center than GR. We plot the radial dependence of the trace of the matter stress-energy tensor $\mathfrak{T}$ for QRG ($\epsilon=-0.01$) in comparison to the GR ($\epsilon=0$) case as shown by Fig.~\ref{fig:TEC1}. Notably, for $\epsilon=+0.01>0$, the maximum compactness $C_\text{max}\simeq 0.68$ can be realized.

For the QRG, we plot $\Delta-C$ and $\mathfrak{T}-C$ phase-spaces at the center as shown by Fig.~\ref{fig:TEC2}. The trace anomaly $\Delta\geq 0$ and $\mathfrak{T}\leq 0$ as long as $C\leq 0.752$ which determines the maximum compactness allowed by the QRG. If future observations verify existence of compact stars with $C>0.752$, it would indicate possibility of positiveness of the matter trace or equivalently negative value of the trace anomaly at the stellar center.  Remarkably, the trace anomaly diverges, $\Delta \to -\infty$, at the stellar center as $C\simeq 0.96$, which coincides with the maximum compactness of a maximally compact EoS, $p_r=v_r^2 (\rho-\rho_s)$ with $v_r=c$, where causality still holds.
\subsection{From macroscopic to microscopic structure}\label{Sec:structure}

Up to this moment, we obtained some important features of the QRG on the macroscopic level of compact stellar structure. The theory modifies the TOV equation by introducing a new repulsive force to the total force budget. This force allows for more stellar mass within same size, whereas the adiabatic sound speed does not violate the conformal upper limit. On the other hand, we confirmed the validity of the TEC, or equivalently the positiveness of the trace anomaly, which sets an upper limit on the compactness $C\simeq 0.752$.  The model is characterized by vanishing of the surface radial pressure, while its surface density is above the nuclear saturation density as obtained by the KB-EoS $p_r=v_r^2(\rho-\rho_s)$ with $v_r \simeq c/\sqrt{3}$, this makes the star a self-bound object. This is also confirmed by the divergence of radial adiabatic index at the stellar surface, which allows for very stable heavy stars $\simeq 2 M_\odot$ even if the sound speed is small. Although the model can describe stable stars macroscopically, one must extend investigations to examine the model on the microscopic level.

In the case of a self-bound state on the microscopic level, it is natural to consider the case of quark star model. It has been conjectured that quark matter could be the true ground state of matter at zero pressure as proposed by \citet{Witten:1984rs}, see also \citep{Farhi:1984qu}. Indeed, deconfinement of quarks process is a strong interaction which converts baryons into two flavour quarks, up ($u$) and down ($d$) quarks. Whether $ud$ plasma has been produced inside a NS core where the density approaches (3$-$5)$\times 10^{14}$ g/cm$^3$, or the NS already contains small fraction of quark matter of cosmological origin; half of $d$ quarks convert into strange ($s$) quarks via equilibrium strangeness weak interaction
\begin{equation}
    u+d \rightarrow u+s.
\end{equation}
The time suggested, depends on the temperature and the EoSs, for this conversion $10\, \text{ms}\lesssim \tau \lesssim 10\,\text{min}$, cf. \cite{Olinto:1987PhLB,Doroba:1989AcPPB,Heiselberg:1993PhRvD}. Even a much shorter supersonic conversion $\tau \sim 0.1\, \text{ms}$ has been suggested, since hydrodynamical instability of subsonic conversion favors fast conversion \cite{Horvath:1988PhLB,Drago:2005yj}. At any case, strange quark matter hypothesis implies a new minimum of the energy per baryon $\mu_b=\rho c^2/n_b$, at zero pressure, lower than the energy per baryon of Iron nuclei $^{56}$Fe ($ < 930~\text{MeV}$). In a compact stellar objects, the quark chemical potential is $\mu \simeq 500~\text{MeV}$ at most and the temperature is negligible (zero in practice). We consider quark matter with three $uds$ flavors, whereas the quark masses are assumed to be negligible with respect to the chemical potential, which is acceptable approximation in the present context. In this simplest case, one could adopt an MIT bag model in which quarks are confined to a bag, $B$, of a QCD vacuum, which have been introduced as a phenomenological constant to compensate the energy difference between confined ordinary nuclear matter and deconfined quarks\footnote{The bag constant is measured in MeV$^4$ in the natural units system (i.e. $\hbar=c=1$), whereas it is measured in MeV/fm$^3$ in nuclear physics units, which will be used in this section. We remark that $B_\text{natural}=B_\text{nuclear}(\hbar c)^3$.}. The energy density, for massless and non-interacting quark, is given by
\begin{eqnarray}
    \rho c^2= \rho_q c^2 + B,
\end{eqnarray}
where the microscopic quark kinetic energy, $\rho_q c^2$, of an ultra-relativistic Fermi gas combined with the perfect flavor-color symmetry (3 color, 3 flavour and 2 spin), is given by \cite{2007ASSL..326.....H}
\begin{equation}
    \rho_q c^2= \frac{9}{4}\pi^{2/3} \hbar c \, n_b^{4/3},
\end{equation}
where $\hbar$ is the reduced Planck’s constant and $n_b$ is the baryon density number. Consequently, the microscopic pressure
\begin{equation}
    p_r=n_b^2 \frac{d}{dn_b}\left(\frac{\rho c^2}{n_b}\right)=\frac{3}{4}\pi^{2/3} \hbar c \, n_b^{4/3}-B.
\end{equation}
At the stellar surface ($p_r=0$), the baryon density number
\begin{equation}
    n_b=\frac{2\sqrt{2}}{\sqrt{\pi}}\left(\frac{B}{3\hbar c}\right)^{3/4},
\end{equation}
which consequently implies a self-bound state at zero pressure. This simple case can be described by the MIT bag model EoS
\begin{equation}\label{eq:MIT-EoS2}
    p_r=\frac{1}{3}(\rho c^2-4B).
\end{equation}
Clearly, the bag constant $B$ is related to the surface density (at which the radial pressure vanishes) by $\rho_\text{s}=4B/c^2$. In order to investigate the similarities between QRG and the MIT bag model, we rewrite the field equations \eqref{eq:Feqs} in the form of
\begin{equation}
    \tilde{\rho} c^2=\rho c^2+\widetilde{B}, \, \tilde{p}_r=p_{r}-\widetilde{B} \,\text{and}\, \tilde{p}_t=p_{t}-\widetilde{B},
\end{equation}
where $\widetilde{B}=2{\epsilon  \kappa \ell^2} \left[\tilde{\rho} c^2 (\tilde{p}_r+2\tilde{p}_t)-\tilde{p}_t(\tilde{p}_t+2\tilde{p}_r)\right]$. This leads to the following relation
\begin{equation}\label{eq:eff-EoS}
    \tilde{p}_r=\frac{1}{3}(\tilde{\rho}c^2-4 B_{eff}),
\end{equation}
where
\begin{eqnarray}\label{eq:eff-bag}
B_{eff}&=&B+\widetilde{B}=B+2 \epsilon  \kappa \ell^2 \left[\tilde{\rho} c^2 (\tilde{p}_r+2\tilde{p}_t)-\tilde{p}_t(\tilde{p}_t+2\tilde{p}_r)\right].
\end{eqnarray}
At the GR limit, $\epsilon \to 0$, we obtain $B_{eff}=B$. Equation \eqref{eq:eff-EoS} shows the role of the nonminimal coupling between matter and geometry due to QRG to effectively mimic bag constant similar to the cases when QCD corrections are included.\\

It remains to question the validity of the QRG provided in this study on the microscopic level in comparison to the physical constraints of the MIT bag model. We note that the strange quark star would be valid, where at the stellar surface, the following constraint is fulfilled
\begin{equation}
    \mu_b=\frac{\rho c^2}{n_b}=\sqrt{2\pi}(3\hbar c)^{3/4} B^{1/4}<930\,\text{MeV}.
\end{equation}
This simply requires an upper bound to the bag constant  $B<91.34$ MeV/fm$^3$. On the other hand, for nuclei to be stable with respect to the $ud$ matter $B>56.64$ MeV/fm$^3$. A more restrictive lower bound on the bag constant is by requiring that the neutrons should not coagulate into droplets of the $ud$ matter, which leads to $B>58.92$ MeV/fm$^3$, c.f. \cite{2007ASSL..326.....H}. Fortunately, in our case, the bag constant $B=\rho_s c^2/4$ in KB EoS, namely \eqref{eq:EoS}, is completely determined by the model parameter, as for example for the pulsar PSR J0740+6620 we have $\rho_\text{s}=-(c_0/c_1)\rho_\star \approx 4.32\times 10^{14}$~g/cm$^3$, as obtained in the previous section, which directly gives
\begin{equation}
    B=-\frac{1}{4}(c_0/c_1)\rho_\star c^2 \simeq 60.58\, \text{MeV/fm}^3.
\end{equation}
This result confirms the validity of the QRG with the microscopic strange quark star hypothesis.

We Recall the results of Table \ref{Tab:neg_para}, which obtain estimated radii of the pulsar PSR J0740+6620 for different values of $\epsilon$ assuming that the conjectured conformal upper limit on the sound speed is valid. Therefore, it is straightforward to see that $B=32.67$ MeV/fm$^3$ for $\epsilon=0$ and $B=21.08$ MeV/fm$^3$ for $\epsilon=+0.01$ much less than the lower bound, which means that deconfined quarks would be visible in the atomic nuclei. This provides another evidence to exclude the GR ($\epsilon=0$) and QRG with $\epsilon>0$ cases, but from microscopic physics.


\section{Estimation of PSR J0952–0607 Radius}\label{Sec:J0952–0607}


A recent observation, using multicolor light curves, measures the heaviest pulsar ever-seen  PSR J0952\textendash{0607}  with mass $M = 2.35\pm 0.17\, M_\odot$ \cite{Romani:2022jhd}. The rotational frequency of this ms pulsar is 707 Hz. Since its mass near to $M_\text{TOV}$, the size measure of this pulsar provides an important piece of information to understand the nature at such dense matters. This is the main aim of the present section.
\subsection{Conformal sound speed constraint on the radius}\label{Sec:radius}
We assume the conformal constraint on the upper bound of the sound speed is applied everywhere inside the pulsar PSR J0952–0607. Therefore, we take the possible maximum value at the stellar center, i.e.
\begin{equation}
    \lim_{r\to 0} v_r^2/c^2 = 1/3.
\end{equation}
Using field equations \eqref{eq:Feqs3}, or simply by realizing the density coefficient $c_1$ in the KB--EoS \eqref{eq:EoS_const}, we obtain the following constraint
\begin{equation}\label{eq:conf_cs}
    {a_2(a_0+a_2)\over 5a_2^2-\left(32a_0^3-168a_0^2 a_2+260a_0a_2^2-80a_2^3\right)\epsilon}= 1/3.
\end{equation}
Inserting \eqref{eq:const} into the above inequality and by taking the nonminimal coupling parameter $\epsilon=-0.01$ as constrained by the astrophysical observations of the pulsar PSR J0740+6620, we calculate the compactness parameter of the pulsar PSR J0952–0607 as $C = 0.53$. Consequently, the corresponding KB model parameters read $a_0 = 0.56$, $a_1 = -1.31$, $a_2 = 0.75$. Given its observed mass $M=2.35\pm0.17 M_\odot$, we obtain the corresponding radius $R=13.21\pm 0.96$ km. This result is in agreement with radii measured values of compact stars with relevant masses, see Table \ref{Tab:obs_data}.

We accordingly provide some useful physical quantities. At the center $\rho(0) = 6.99\times 10^{14}$ [g/cm$^3$] $\approx 2.59 \rho_\text{nuc}$, $p_r(0) = p_t(0)= 9.18\times 10^{34}$ [dyn/cm$^2$] and $Z(0)=0.92$. At the surface $\rho(R) = 3.80\times 10^{14}$ [g/cm$^3$] $\approx 1.41 \rho_\text{nuc}$, $p_r(R) = 0$ [dyn/cm$^2$], $p_t(R) = 3.38\times 10^{34}$ [dyn/cm$^2$] and $Z(R)=0.45$. Similar to PSR J0740+6620, we evaluate the KB EoSs of the pulsar PSR J0952–0607
\begin{eqnarray}\label{eq:J0952KBEoS}
    p_r&=&0.333 c^2 (\rho-3.92\times 10^{14}) [\text{dyn/cm}^2],\nonumber\\
    p_t&=&0.214 c^2 (\rho-2.22\times 10^{14}) [\text{dyn/cm}^2].
\end{eqnarray}
In addition, we obtain the best fit EoSs
\begin{eqnarray}\label{eq:J0952bestEoS}
    p_r&=&0.310 c^2 (\rho-3.69\times 10^{14}) [\text{dyn/cm}^2],\nonumber\\
    p_t&=&0.194 c^2 (\rho-2.08\times 10^{14}) [\text{dyn/cm}^2].
\end{eqnarray}
Clearly both are in a perfect agreement. In the present model, for the pulsar PSR J0952\textendash{0607}, we have $c_1 \simeq 1/3$, $c_0\simeq -0.43$ and $\rho_\star\simeq 151.83$ MeV/fm$^3$. This gives $B=-\frac{1}{4}(c_0/c_1)\rho_\star c^2 \simeq 58.97$ MeV/fm$^3$ just above the lower bound. This additionally verifies the validity of the QRG on the stellar microscopic structure.

If we apply same procedure within GR ($\epsilon=0$), by applying the conformal constraint on the sound speed upper limit at the center of the pulsar PSR J0952–0607, then we obtain the compactness $C=0.42$ with a corresponding radius $R=16.41\pm 1.19$ km, which is in a strong tension with the measured radii values of relevant stellar masses. On the contrary, if we restrict the radius to acceptable range $\sim 13$ km, one can not hold the sound speed at its conformal upper bound. We, additionally, determine the KB model parameters are $a_0 = 0.37$, $a_1 = -0.92$, $a_2 = 0.55$. At the stellar center, we obtain $\rho(0) = 3.29\times 10^{14}$ [g/cm$^3$] $\approx 1.22 \rho_\text{nuc}$, $p_r(0) = p_t(0)= 3.28\times 10^{34}$ [dyn/cm$^2$] and $Z(0)=0.58$. At the surface, we obtain $\rho(R) = 2.11\times 10^{14}$ [g/cm$^3$] $\approx 0.78 \rho_\text{nuc}$, $p_r(R) = 0$ [dyn/cm$^2$] and $p_t(R) = 1.20\times 10^{34}$ [dyn/cm$^2$] and $Z(R)=0.32$. This in turn determines the bag constant $B=\frac{1}{4}\rho_s c^2 \simeq 30.72$ MeV/fm$^3$ below the physical lower bound. Much lower values are obtained for QRG with $\epsilon>0$. We interpret this result by the presence of an additional collapsing force $F_\text{QRG}<0$ (or it vanishes in the GR case), which requires larger stellar size (less dense medium) for the conformal upper limit on the sound speed to be hold. This finally leads to smaller surface density ($\rho_\text{s}=4B/c^2$) violating the lower bound on the bag constant. This is unlike the QRG with $\epsilon<0$ which induces an additional repulsive force allowing for a stable stellar configuration at smaller radii (more dense medium) compatible with quark matter physical constraints. The present results show the novel role of the matter-geometry nonminimal coupling within the QRG to obtain a stellar model consistent with macroscopic and microscopic physical constraints.
\subsection{Mass-radius diagram and crustal structure}\label{Sec:M-R}
One of the characterizing features which distinguishes strange quark star from normal matter star is the mass-radius relation at small masses, $M\lesssim 0.3 M_\odot$, where the mass decreases monotonically as $M\propto R^3$. This is radically different from normal matter where mass in fact decreases as radius increases at relevant low masses. The low-mass strange matter core is dominated by the confining QCD vacuum, whereas gravitational force is neglected and quark matter is highly incompressible. In this case, the density is nearly constant and Newtonian gravity can be applied. Therefore, a mass-radius diagram would be a useful tool to examine the viability of the model with different observational datasets with a variety of masses.

\begin{figure*}
\centering
\includegraphics[scale=0.5]{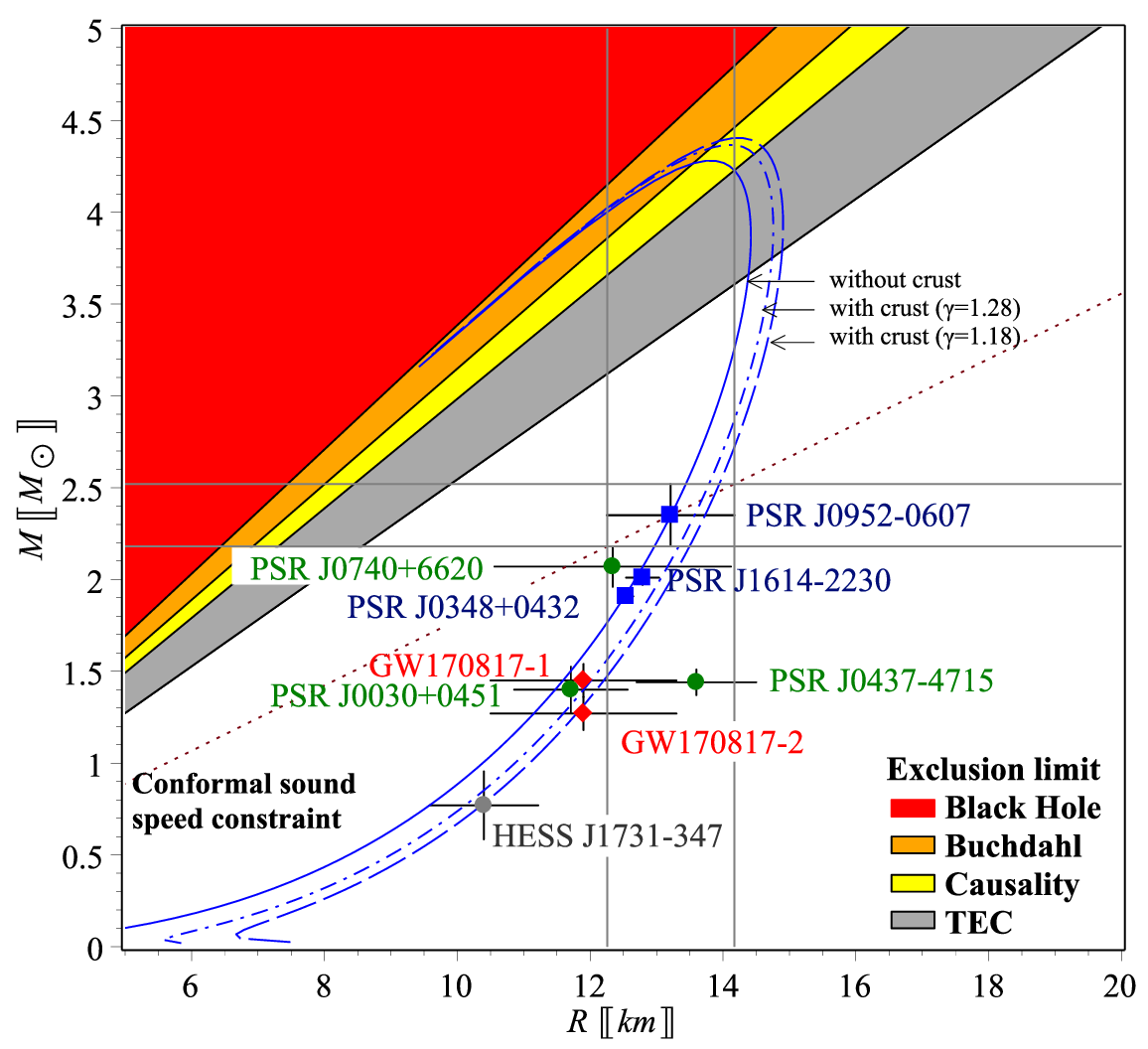}
\caption{Mass-Radius diagram: We use $\epsilon=-0.01$ and $\rho_s =3.8 \times 10^{14}$ g/cm$^3$. The colored regions represent excluded values according to several physical constraints. The diagonal dotted line represents the conformal constraint on the sound speed $v_r^2/c^2=1/3$. We present several observational data as provided by Table \ref{Tab:obs_data} including the lightest pulsar HESS J1731\textendash{347} and the heaviest J0740+6620 in addition to our estimated radius value of the pulsar J0952\textendash{0607}. Clearly the conformal sound speed limit is hold for all pulsars. In the low mass range $M \propto R^3$ which characterizes strange quark star models. For a bare quark star ignoring the crust contribution the radius is underestimated by $<$ 1 km. However, the crust contribution is provided according to equation \eqref{eq:crust} for two different values of adiabatic index $\gamma=1.28$ and $\gamma=1.18$.}
\label{Fig:MR_diag}
\end{figure*}

In our case, we solve the density profile \eqref{eq:Feqs3}, for radius at compactness values $0\leq C \leq 1$, assuming that the surface density $\rho_s=3.8\times 10^{14}$ g/cm$^3$ relevant to the obtained value for the pulsar PSR J0952–0607 via KB EoS \eqref{eq:J0952KBEoS}.  Then, we plot the corresponding mass\textendash{radius} diagram as shown in Fig. \ref{Fig:MR_diag} (without crust contribution), which is a typical strange quark matter EoS pattern. The figure includes observational constraints of nine pulsars, as provided by Table \ref{Tab:obs_data}, from the lightest pulsar HESS J1731\textendash{347} (gray solid circle) to the heaviest J0952\textendash{0607} (blue solid box) in addition to other observational constraints on the intermediate mass range from NICER and LIGO/Virgo collaboration. We note that the radii of the pulsars J0952\textendash{0607}, J0348+0432 and J1614\textendash{2230} are obtained by assuming that the conjectured conformal upper bound on the sound speed is applied at the center of the pulsar. Those are indicated by blue solid boxes in Fig. \ref{Fig:MR_diag}. For NICER measurements of the pulsar J0030+0451 we include the latest analysis of updated NICER + XMM-Newton data (using ST+PDT model) \citep{Vinciguerra:2023qxq}. We dropped other older analysis of NICER data of the pulsar J0030+0451 \cite{Miller:2019cac,Raaijmakers:2019qny}, since they are in tension with GW170817. All other NICER measurements of the pulsars J0740+6620 and J0437–4715 are indicated by green solid circles whereas gravitational wave signals are indicated by red solid diamonds. In general, the mass-radius curve are in agreement  at 68\% confidence interval with observational data except for the pulsar J0437–4715 case which indeed has relatively large size \cite{Gonzalez-Caniulef:2019wzi}. If upcoming analysis of this pulsar obtains it radius of $R\sim 12$ km, the present model would be in agreement. More pulsars’ observational data by different observational techniques are given in appendix \ref{App:pulsars} as provided by Table \ref{Tab:pulsar_list} with a supplementary numerical estimates of some important physical features for those pulsars, see Table \ref{Tab:phys_features}.
\begin{table}[h]
    \centering
    \caption{Astrophysical observations of mass and radius of some pulsars.}
    \begin{ruledtabular}
    \begin{tabular}{lccc}
       Pulsar                  & Mass ($M_\odot$)       & Radius (km)             & Ref.\\[5pt]
\hline
       J0952\textendash{0607}  & $2.35\pm 0.17$         & $13.21 \pm 0.96$\footnote{\label{note1}The radius value is predicted by the present model assuming that the conformal constraint on the sound speed $v_r^2/c^2\lesssim 1/3$ is hold at the stellar core.}     & \cite{Romani:2022jhd}\\[5pt]
       J0740+6620              & $2.07 \pm 0.11$        & $12.34^{+1.89}_{-1.67}$              & \cite{Legred:2021hdx}\\[5pt]
       J0348+0432              & $2.01 \pm 0.04$        & $12.79 \pm 0.25^\text{\ref{note1}}$        & \citep{Antoniadis:2013pzd}\\[5pt]
       J1614\textendash{2230}  & $1.908 \pm 0.016$      & $12.54 \pm 0.11^\text{\ref{note1}}$        & \citep{Demorest:2010bx,Fonseca:2016tux,NANOGRAV:2018hou}\\[5pt]
       J0030+0451\footnote{We use the result of ST+PDT model which estimates radii in line with current observations \citep{Vinciguerra:2023qxq}.}  & $1.40^{+0.13}_{-0.12}$ & $11.71^{+0.88}_{-0.83}$ & \citep{Vinciguerra:2023qxq}\\[5pt]
       J0437\textendash{4715}  & $1.44 \pm 0.07$        & $13.6^{+0.9}_{-0.8}$          & \citep{Reardon:2015kba,Gonzalez-Caniulef:2019wzi}\\[5pt]
       GW170817-1              & $1.45 \pm 0.09$        & $11.9 \pm 1.4$          & \citep{LIGOScientific:2018cki}\\[5pt]
       GW170817-2              & $1.27 \pm 0.09$        & $11.9 \pm 1.4$          & \citep{LIGOScientific:2018cki}\\[5pt]
       HESS J1731\textendash{347}   & $0.77_{-0.17}^{+0.20}$ & $10.4^{+0.86}_{-0.78}$  & \citep{2022NatAs...6.1444D}\\[5pt]
    \end{tabular}
    \end{ruledtabular}
        \label{Tab:obs_data}
\end{table}

We briefly discussed the possibility for a NS to have a small seed of quark matter of cosmological origin or by forming it at the NS center in Sec. \ref{Sec:structure}. Since there is no Coulomb barrier in the stellar interior, the weak interaction $u+d \to u+s$ consumes free neutrons converting the star entirely to quark matter. In this case, bare strange star is possible since the surface density is $10^3$ larger than neutron drip density ($\simeq 10^{11}$ g/cm$^3$). However, thin and light crustal structure of normal matter is also possible where free neutrons at the surface no longer exist, and there is a sufficiently large Coulomb barrier between crust normal matter and the interior quark matter \citep{Zdunik:2001yz}. The development of normal crust matter could be a NS remnant or via accretion. At any case, it would have a crustal structure with density at neutron drip density $\simeq 4\times 10^{11}$ g/cm$^{3}$ at the bottom of the crust where the light and thin crust approximation is most likely. Let $\rho_\text{cr}$ and $p_\text{cr}$ be the density and pressure at the crust bottom, while the average adiabatic index of the otter crust be $<\gamma>$. Assuming that the solid crust consists of dense matter below the neutron drip which can be described by BPS model \citep{Baym:ApJ1971}, then the crust mass and thickness are given by \citep{Zdunik:2001yz},
\begin{eqnarray}\label{eq:crust}
    \Delta M_\text{cr}^\text{stat}&=&8\pi R^3 \frac{(1-C)}{C}\frac{p_\text{cr}}{c^2},\\[5pt]
    \Delta R_\text{cr}^\text{stat}&=&\frac{2<\gamma>}{<\gamma>-1}\frac{R}{C}\frac{p_\text{cr}}{\rho_\text{cr}c^2}.
\end{eqnarray}
Taking the crust density $\rho_\text{cr}=4.3\times 10^{11}$ g/cm$^{3}$ equal to the neutron drip density and the crust pressure $p_\text{cr}=7.8\times 10^{29}$ dyn/cm$^{2}$ with an average adiabatic index $<\gamma>=1.28$ of normal matter. In general, the crust mass is negligible and it is thickness varies according to the interior mass as shown in Fig. \ref{Fig:MR_diag}. For example, the crust in the case of the massive pulsar J0740+6620 has a negligible mass contribution $\Delta M_\text{cr}^\text{stat}\approx 2.1\times 10^{-5}\, M_\odot$ and a thickness $\Delta R_\text{cr}^\text{stat}\approx 0.46$ km, while in the case of the light pulsar HESS J1731\textendash{347} the crust mass $\Delta M_\text{cr}^\text{stat}\approx 4.41\times 10^{-5}\, M_\odot$ is still tiny with a relatively larger thickness $\Delta R_\text{cr}^\text{stat}\approx 0.87$ km. Clearly, when the mass is below $\sim 0.1M_\odot$ the thickness significantly changes toward much larger values as obtained by Fig. \ref{Fig:MR_diag}.


\section{Summary and Conclusion}\label{Sec:summary}


Neutron stars uniquely provide natural laboratory to test physical conditions beyond terrestrial-based experiments capability. Since matter is so dense and the spacetime curvature is efficient in the stellar interior, modification of general relativity such as  matter-geometry nonminimal coupling would be significant. In the present paper, we provide an extension to GR by including Rastall-like nonminimal coupling but proportional to the gradient of quadratic curvature invariants. Those are similar to the conformal trace anomaly when quantum fields backreaction on the curved spacetime is included.

Recent independent observations which measure mass and radius of pulsars by LIGO/Virgo and NICER provide unprecedented accurate measurements of those quantities, which in turn helps not only to understand the matter nature at high pressure and density conditions and possible EoS but also to constrain the parameter space of suggested modified gravity. Existence of massive NSs $\gtrsim 2 M_\odot$ requires stiff EoS with $v_r^2/c^2 \sim 0.7$, on the contrary to tidal deformability of gravitational wave signals which requires soft EoS. These evidences challenge theoretical models which in general assume perfect fluid with hadron EoS when GR is applied.

In the present work, we assume static spherical symmetric spacetime, where the geometric sector of the stellar interior is governed by KB metric potentials \eqref{eq:KB}. The fluid is anisotropic \eqref{Tmn-anisotropy}, which is more realistic in practice, where the anisotropic force becomes repulsive if $p_t>p_r$ (strong anisotropy condition). The general relativistic minimal coupling procedure is relaxed and replaced by QRG which includes quadratic nonminimal coupling \eqref{eq:QRastall}. The QRG field equations are given by equations \eqref{eq:QRT}.

Accordingly we obtain the density and pressure profiles, namely equations \eqref{eq:Feqs3}, by applying the QRG to the above mentioned configuration of the stellar structure model. The model parameters can be entirely written in terms of the compactness as in the GR case in addition to another dimensionless parameter $\epsilon$ which characterizes the nonminimal coupling deviation from GR (in which $\epsilon=0$). Unlike Rastall gravity the QRG coupling constant is identical to Einstein's coupling, since the field equations in GR and QRG are identical in the weak field limit, see appendix \ref{App:WFL}. Also, both theories are identical in vacuum, since $\mathfrak{R}=0=\mathfrak{R}_{\mu\nu}$. In effect Schwarzschild vacuum solution is valid within QRG framework. However, for nondegenerate $\mathfrak{R}_{\mu\nu}$, another vacuum state can be obtained in QRG where $\mathfrak{R}=-1/3\epsilon\ell^2$, which has been omitted in the present study as $\epsilon=0$ is not allowed.

We use the Observational data constraints by NICER+XMM-Newton on the pulsar PSR J0740+6620 to determine the parameter $\epsilon \sim -0.01$. In addition, if the conformal upper bound on the sound speed is applied at the center, we show that $\epsilon\geq 0$ values are excluded by NICER at $\gtrsim 1.6\sigma$ level, see Table \ref{Tab:neg_para}. This may indicate an underlying similarities between QRG and conformal anomaly which requires a necessarily negative coefficient. For the pulsar PSR J0740+6620, NICER constraints confirm that the adiabatic radial sound speed does not violate the conformal upper bound $v_r^2/c^2=1/3$, if QRG is assumed unlike GR. The trace anomaly is positive which confirms the negative value of the matter trace inside the star and indicates that the conformal symmetry is broken inside the star. This in turn sets an upper bound on the maximum compactness $C_\text{max}=0.752$ which is $4\%$ higher than GR.

The analysis of the contributing forces of the hydrostatic equilibrium confirms that, in addition to the repulsive hydrostatic and anisotropic (strong anisotropy condition) forces, there is an additional repulsive force due to nonminimal coupling with $\epsilon<0$. This regulating force contribute to keep the sound speed below the conformal constraint unlike the GR case which requires sound speed above this limit. The divergence of the radial adiabatic index at the star surface identifies that the star is self-bound. This feature together with soft EoS (i.e. $v_r^2/c^2 \sim 1/3$) suggests that the star to be a quark star rather a NS with normal matter.

We show that the KB ansatz effectively relates the density and pressures linearly up to constants, which we call them KB-EoSs \eqref{eq:EoS}. The investigation on the microscopic structure shows that the nonminimal coupling contributes to the matter density and pressure effectively as a bag constant of the MIT quark star model. By requiring the conformal sound speed constraint inside the pulsar PSR J0740+6620 to be applied, the estimated bag constant is consistent with the physical range $58.92 < B=60.58$ MeV/fm$^{-3}< 91.34$ for $\epsilon=-0.01$. However, the values $\epsilon \geq 0$ are excluded since they provide small $B$ below the allowed value.

Since the radius of the pulsar PSR J0952–0607 is not yet measured, we apply the conformal constraint on the sound speed to estimate its radius according to the QRG correction. Our estimated value, for $\epsilon=-0.01$, is $R=13.21\pm 0.96$ km in agreement with the measured radii of other pulsars with comparable masses. On the microscopic level we confirm the validity of the numerical values of the model parameters with physical constraints from MIT bag model. Our calculations show that same procedure can be applied in GR but the estimated radius then is $R=16.41 \pm 1.19$ km which is incompatible with other pulsars within the same mass range, whereas the case gets worse for $\epsilon>0$ values. This in turn estimates lower surface density which is related to small bag constant values not valid within its physical range.

We provide a mass-radius diagram corresponding to the suggested QRG with $\epsilon =-0.01$ and surface density $\rho_s=3.8 \times 10^{14}$ g/cm$^3$ above $\rho_\text{nuc}$. The diagram fits well with wide range of observational data including NICER and LIGO/Vigo collaboration. In conclusion, within QRG same physics can be applied successfully to low-mass, HESS J1731\textendash{347}, and high-mass, PSR J0740+6620, compact stars. Also we confirm that nonminimal coupling can provide a better framework to keep the sound speed below the conformal upper limit, while macro and micro physical constraints are satisfactory fulfilled.

\begin{acknowledgments}
The author would like to thank A. Awad and A. Golovnev for helpful comments during the preparation of this paper.
\end{acknowledgments}

\appendix
\section{The Classical Limit}\label{App:WFL}
In this section, we investigate the classical limit of the QRG field equations, where the gravitational fields are assumed to be weak and the relativistic effects do not contribute due to low speed approximation as in Poisson's equation. In order to simplify the presentation, we include only $\mathfrak{R}^2$ term in equation \eqref{eq:QRastall}, i.e.
\begin{equation}
     \nabla_{\alpha}\mathfrak{T}{^\alpha}{_\beta}=-\frac{\epsilon \ell^2}{\kappa}\, \partial_{\beta} \mathfrak{R}^2.
\end{equation}
However, the conclusion will be the same for the full form. Then, equation \eqref{eq:eff_Tab} reads
\begin{equation}
\widetilde{\mathfrak{T}}{^\alpha}{_\beta}=\mathfrak{T}{^\alpha}{_\beta}+\frac{\epsilon \ell^2}{\kappa} \delta^\alpha_\beta\mathfrak{R}^2.
\end{equation}
Using $\mathfrak{R}=-\kappa \widetilde{\mathfrak{T}}$, the contraction of the above gives
\begin{equation}
    \widetilde{\mathfrak{T}}={1 \pm \sqrt{1-16 \epsilon \kappa \ell^2 \mathfrak{T}} \over 8 \epsilon \kappa \ell^2}.
\end{equation}
At the linear order, we have $\sqrt{1-16 \epsilon \kappa \ell^2 \mathfrak{T}} \approx 1- 8 \epsilon \kappa \ell^2 \mathfrak{T}$, where $|16 \epsilon \kappa \ell^2 \mathfrak{T}| \ll 1$. This proves that $\widetilde{\mathfrak{T}} \approx \mathfrak{T}$ or $\widetilde{\mathfrak{T}} \approx \frac{1}{4\epsilon \ell^2 \kappa}-\mathfrak{T}$ at the weak field limit. Similar to the full form in Sec. \ref{Sec:QRgrav} we drop the latter solution which allows vacuum solution $\mathfrak{R}=-\frac{1}{4\epsilon \ell^2}$ in the present case, since $\epsilon= 0$ is not valid. Consequently, for the first solution, Ricci tensor in \eqref{eq:Ricci_tensor} follows the relation
\begin{equation}
    \mathfrak{R}_{\alpha\beta}=\kappa\left(\mathfrak{T}_{\alpha\beta}-\frac{1}{2}g_{\alpha\beta}\mathfrak{T}\right),
\end{equation}
which is identical to the GR theory. For the $tt$-component, we write
\begin{equation}
    \mathfrak{R}_{tt}=\nabla^2 \psi=\frac{\kappa}{2} \rho,
\end{equation}
where $\psi$ denotes the conventional Newtonian potential. The comparison with Poisson equation, $\nabla^2 \psi=\frac{4\pi G}{c^4} \rho$, allows us to obtain the QRG coupling constant
\begin{equation}
    \kappa=\frac{8\pi G}{c^4}=\kappa_E,
\end{equation}
identical to GR. Similarly, one would obtain the same result if the full QRG, namely equation \eqref{eq:QRastall}, is applied. In conclusion, QRG does not modify the coupling constant unlike Rastall gravity, where Rastall's correction ($\propto \mathfrak{R}$) spoils up Einstein tensor and finally leads to a modified coupling constant \eqref{eq:Rconst}.
\section{More pulsars' observational data}\label{App:pulsars}
In this section, we extend our study to include more pulsars' observational data. This is to test the viability of the QRG scenario with larger dataset of different observational approaches. In Table \ref{Tab:pulsar_list}, we give a list of four categories of pulsars according to their types and observational technique. As noted before that the radii measurements are not an easy task, so some of the presented values of radii are in fact model dependent in way or another. We generated this table, for $\epsilon=-0.01$, by imposing the conformal constraint on the sound speed to derive the corresponding pulsar radius (if not observed) given its mass at 68\% CL. Those predicted values will be labeled by ``p" letter in Table \ref{Tab:pulsar_list}. Consequently, we list the corresponding compactness and the KB parameters \eqref{eq:const}. In addition, we provide the estimated mean mass as given by equation \eqref{eq:Mass} to ensure its agreement with the observed value within 68\% CL. We note that since $C<C_\text{max}=0.752$, the pulsar is stable and verifies all physical conditions.

\begin{table}[h!]
\caption{The KB model parameters (with $\epsilon=-0.01$) of different types of observed pulsars which generate mass and compactness in agreement with the observed mass and radius values within 1$\sigma$ CL.}
\label{Tab:pulsar_list}
\renewcommand{\baselinestretch}{1}
\footnotesize{
\begin{ruledtabular}
\begin{tabular*}{\textwidth}{@{\extracolsep{\fill}}llccccccc@{}}
\multicolumn{1}{c}{Pulsar} & \multicolumn{1}{c}{Ref} & observed mass &  radius & estimated mass & \multicolumn{1}{c}{$C$}& \multicolumn{1}{c}{$a_0$}
& \multicolumn{1}{c}{$a_1$}& \multicolumn{1}{c}{$a_2$}\\
   & &  ($M_{\odot}$) &   [{km}] &  ($M_{\odot}$) &      &     &  &    \\
\hline
\multicolumn{9}{c}{High-mass X-ray binaries}\\
\hline
Her X\textendash{1}         &\cite{Abubekerov_2008}        &  $0.85\pm 0.15$    &  $7.56\pm 1.33^\text{p}$   & $0.82$& $0.29$ &  $0.21$    & $-0.56$    & $0.35$     \\
4U 1538\textendash{52}      &\cite{Rawls:2011jw}        &  $0.87\pm 0.07$    &  $7.74\pm 0.62^\text{p}$ & $0.84$ & $0.30$&  $0.21$    & $-0.57$    & $0.36$     \\
SMC X\textendash{1}         &\cite{Rawls:2011jw}        &  $1.04\pm 0.09$    &  $8.34\pm 0.72^\text{p}$   & $1.00$& $0.34$ &  $0.26$    & $-0.67$    & $0.41$     \\
LMC X\textendash{4}         &\cite{Rawls:2011jw}           &  $1.29\pm 0.05$    &  $9.10\pm 0.35^\text{p}$  & $1.27$ & $0.40$ &  $0.34$    & $-0.85$    & $0.51$     \\
Cen X\textendash{3}         &\cite{Rawls:2011jw}            &  $1.49\pm 0.08$    &  $9.48\pm 0.51^\text{p}$ & $1.47$ & $0.44$ &  $0.40$    & $-0.99$    & $0.59$     \\
Vela X\textendash{1}         &\cite{Rawls:2011jw}            &  $1.77\pm 0.08$    &  $9.97\pm 0.45^\text{p}$ & $1.75$ &  $0.50$ & $0.51$     & $-1.20$    & $0.70$     \\
\hline
\multicolumn{9}{c}{Low-mass X-ray binaries (thermonuclear bursts/quiescence)}\\
\hline
4U 1608-52      &\cite{Ozel:2016oaf}           &  $1.57^{+0.30}_{-0.29}$     &  $9.99\pm 1.91^\text{p}$    & $1.58$ & $0.46$ &  $0.44$    & $-1.06$    & $0.62$     \\[3pt]
KS 1731-260     &\cite{Ozel:2016oaf}           &  $1.61^{+0.35}_{-0.37}$    &  $10.06\pm 2.31^\text{p}$     & $1.62$ & $0.46$ &  $0.45$    & $-1.09$    & $0.64$     \\[3pt]
EXO 1745-248    &\cite{Ozel:2016oaf}           &  $1.65^{+0.21}_{-0.31}$      &  $10.50\pm 1.59^\text{p}$  & $1.66$ & $0.46$ &  $0.44$    & $-1.06$    & $0.62$     \\[3pt]
4U 1820-30      &\cite{Ozel:2016oaf}           &  $1.77^{+0.25}_{-0.28}$     &  $11.26\pm 1.59^\text{p}$   & $1.78$ & $0.49$ &  $0.44$    & $-1.06$    & $0.62$     \\[3pt]
4U 1724-207     &\cite{Ozel:2016oaf}           &  $1.81^{+0.25}_{-0.37}$    &  $11.52\pm 1.72^\text{p}$   & $1.82$ & $0.46$ &  $0.44$    & $-1.06$    & $0.62$     \\[3pt]
SAX J1748.9-2021&\cite{Ozel:2016oaf}           &  $1.81^{+0.25}_{-0.37}$     &  $11.52\pm 1.91^\text{p}$   & $1.82$ & $0.46$ &  $0.44$    & $-1.06$    & $0.62$     \\[3pt]
M13             &\cite{Webb:2007tc}            &  $1.38^{+0.08}_{-0.23}$     &  $9.95^{+0.24}_{-0.27}$  & $1.39$ & $0.41$ &  $0.35$    & $-0.88$    & $0.53$     \\[3pt]
X7              &\cite{Bogdanov:2016nle}       &  $1.4$     &  $11.1^{+0.8}_{-0.7}$    & $1.41$ & $0.37$ &  $0.30$    & $-0.76$    & $0.47$     \\
\hline
\multicolumn{9}{c}{Millisecond Pulsars}\\
\hline
PSR J0030+0451  &\cite{Raaijmakers:2019qny}    &  $1.34\pm 0.16$    &  $12.71\pm 1.19$ & $1.490$ & $0.381$ &  $0.309$    & $-0.788$    & $0.479$     \\
                &\cite{Miller:2019cac}         &  $1.44\pm 0.16$    &  $13.02\pm 1.24$ & $1.591$ & $0.397$ &  $0.331$    & $-0.838$    & $0.506$     \\
                & \cite{Vinciguerra:2023qxq}    &  $1.40^{+0.13}_{-0.12}$    &  $11.71^{+0.88}_{-0.83}$ & $1.494$ & $0.395$ &  $0.329$    & $-0.832$    & $0.503$     \\[5pt]
PSR J0437-4715  &\cite{Reardon:2015kba,Gonzalez-Caniulef:2019wzi}        &  $1.44\pm 0.07$    &  $13.6\pm 0.9$   & $1.508$ & $0.350$ &  $0.270$    & $-0.700$    & $0.430$     \\
PSR J1614-2230  &\cite{NANOGrav:2017wvv}       &  $1.908\pm 0.016$  &  $12\pm 1$       & $1.922$ & $0.470$ &  $0.477$    & $-1.081$     & $0.634$     \\
PSR J0348+0432  &\cite{Antoniadis:2013pzd}     &  $2.01\pm 0.04$    &  $12.424\pm 0.517$       & $2.043$ & $0.490$ &  $0.485$    & $-1.158$    & $0.673$     \\[5pt]
PSR J0740+6620  & \cite{Miller:2021qha}     &  $2.07\pm 0.11$    &  $14.3\pm 3.5$       & $2.084$ & $0.515$ &  $0.536$    & $-1.259$    & $0.723$\\
                & \citep{Legred:2021hdx}     &  $2.07\pm 0.11$    &  $12.34\pm 1.75$       & $2.088$ & $0.501$ &  $0.508$    & $-1.203$    & $0.696$\\
\hline
\multicolumn{9}{c}{Gravitational-wave Signals}\\
\hline
GW170817-1      &\cite{LIGOScientific:2018cki} &  $1.45\pm 0.09$    &  $11.9\pm 1.4$   & $1.502$ & $0.394$ &  $0.327$    & $-0.829$    & $0.501$     \\
GW170817-2      &\cite{LIGOScientific:2018cki} &  $1.27\pm 0.09$    &  $11.9\pm 1.4$   & $1.320$ & $0.360$ &  $0.283$    & $-0.730$    & $0.447$
\end{tabular*}
\end{ruledtabular}}
\end{table}

The four categories in Table \ref{Tab:pulsar_list} are: (i) High-mass X-ray binaries category, it includes six eclipsing X-ray pulsars in binary systems with high mass companions, where the X-rays from the pulsar in orbital motion are partly blocked by the companions. We note that there are two different mass measurements of the pulsar Her X\textendash{1}, $M=0.85\pm 0.15 M_\odot$ using radial velocity curve \cite{Abubekerov_2008} and $M=1.07\pm 0.36 M_\odot$ using X-rays \cite{Rawls:2011jw}. In the present study, we take the former mass value which has been obtained by a well justified technique with more precise measurement, and the availability of the corresponding radius measurement $R=8.1\pm 0.41$ km using Roche lobe geometry \cite{Gangopadhyay:2013gha}. In general, the predicted radii are at 68\% agreement with \cite{Gangopadhyay:2013gha}. (ii) Low-mass X-ray binaries category, it includes eight pulsars, mass and radius can be simultaneously measured by spectroscopic analyzing the spectral lines obtained during thermonuclear X-ray bursts or in quiescence \cite{Ozel:2016oaf}. Imposing conformal sound speed bound leads to predicted radii are at 68\% agreement with \cite{Ozel:2015fia}. For M13 and X7, which are quiescence low-mass X-ray binaries, we used the observed radii as given by \cite{Webb:2007tc,Bogdanov:2016nle}. In practice, (i) and (ii) categories give less precise measurements of masses in comparison to ToAs analysis which have been used in the third category. (iii) MSPs category includes five pulsars, in this category some radii are already measured by NICER and some are on going. We include the two independent radius measurements of the pulsar PSR J0030+0451 as obtained by \cite{Raaijmakers:2019qny} and \cite{Miller:2019cac} in addition to its latest updated mass-radius analysis of NICER data \cite{Vinciguerra:2023qxq}. (iv) The gravitational-wave signals category includes two signals labeled as GW170817-1 and GW170817-2 as obtained by LIGO and Virgo collaboration.

\begin{table}
\caption{Estimated values of density, sound speed, trace anomaly, redshift and the bag constant. We note that $B_{60}$ is the value of the bag constant normalized to $B=60$ MeV/fm$^3$, where the allowed range is $0.982<B_{60}<1.53$.}
\label{Tab:phys_features}
\renewcommand{\baselinestretch}{1}
\footnotesize{
\begin{ruledtabular}
\begin{tabular*}{\textwidth}{@{\extracolsep{\fill}}l|cc|cc|cc|cc|cc@{\extracolsep{\fill}}}
\multicolumn{1}{c|}{Pulsar}                              &\multicolumn{2}{c|}{$\rho$ [$\rho_\text{nuc}$]} &   \multicolumn{2}{c|}{$v_r^2/c^2$}  &   \multicolumn{2}{c|}{$\Delta(r)$}  & \multicolumn{2}{c|}{Z(r)} & \multicolumn{1}{c}{$B_{60}$}\\ \cline{2-9}
                                        &\multicolumn{1}{c}{center}            &        surface   &   center        &    surface       &  center          &  surface        &  center     & surface       & {}\\
\hline
\multicolumn{10}{c}{High-mass X-ray binaries}\\
\hline
Her X\textendash{1}            & $3.07$     & $2.31$  &  $0.26$   &  $0.24$     &  $0.27$ & $0.31$ & $0.32$ & $0.19$ & $1.47$  \\
4U 1538\textendash{52}         & $3.13$     & $2.33$  &  $0.26$   &  $0.24$     &  $0.27$ & $0.31$ & $0.33$ & $0.19$ & $1.49$  \\
SMC X\textendash{1}            & $3.26$     & $2.32$  &  $0.27$   &  $0.25$     &  $0.26$ & $0.30$ & $0.40$ & $0.23$ & $1.49$  \\
LMC X\textendash{4}            & $3.54$     & $2.32$  &  $0.29$   &  $0.26$     &  $0.24$ & $0.30$ & $0.53$ & $0.29$ & $1.50$  \\
Cen X\textendash{3}            & $3.77$     & $2.33$  &  $0.30$   &  $0.27$     &  $0.22$ & $0.29$ & $0.64$ & $0.34$ & $1.51$  \\
Vela X\textendash{1}           & $4.05$     & $2.29$  &  $0.32$   &  $0.29$     &  $0.20$ & $0.27$ & $0.83$ & $0.42$ & $1.49$  \\
\hline
\multicolumn{10}{c}{Low-mass X-ray binaries (quiescence/thermonuclear bursts)}\\
\hline
4U 1608-52        &$3.77$     &$2.26$  &  $0.31$   &   $0.28$     &  $0.21$ & $0.28$ & $0.70$ & $0.37$ & $1.47$  \\
KS 1731-260       &$3.81$     &$2.26$  &  $0.31$   &   $0.28$     &  $0.21$ & $0.28$ & $0.73$ & $0.38$ & $1.47$  \\
EXO 1785-248      &$3.41$     &$2.05$  &  $0.31$   &   $0.28$     &  $0.21$ & $0.28$ & $0.70$ & $0.37$ & $1.33$  \\
4U 1820-30        &$2.97$     &$1.78$  &  $0.31$   &   $0.28$     &  $0.21$ & $0.28$ & $0.70$ & $0.37$ & $1.15$  \\
4U 1724-207       &$2.84$     &$1.70$  &  $0.31$   &   $0.28$     &  $0.21$ & $0.28$ & $0.70$ & $0.37$ & $1.10$  \\
SAX J1748.9-2021  &$2.84$     &$1.70$  &  $0.31$   &   $0.28$     &  $0.21$ & $0.28$ & $0.70$ & $0.37$ & $1.10$  \\
M13               &$3.20$     &$2.07$  &  $0.29$   &   $0.26$     &  $0.24$ & $0.29$ & $0.55$ & $0.30$ & $1.34$  \\
X7                &$2.27$     &$1.55$  &  $0.28$   &   $0.25$     &  $0.25$ & $0.30$ & $0.47$ & $0.26$ & $0.99$  \\
\hline
\multicolumn{10}{c}{Millisecond Pulsars}\\
\hline
PSR J0030+0451    &$2.165$     &$1.459$  &  $0.283$   &   $0.254$     &  $0.245$ & $0.298$ & $0.483$ & $0.271$ & $0.921$  \\
                  &$2.190$     &$1.444$  &  $0.288$   &   $0.258$     &  $0.239$ & $0.296$ & $0.520$ & $0.288$ & $0.911$  \\
                  &$2.447$     &$1.618$  &  $0.288$   &   $0.257$     &  $0.240$ & $0.296$ & $0.516$ & $0.286$ & $1.021$  \\[5pt]
PSR J0437-4715    &$1.597$     &$1.121$  &  $0.275$   &   $0.248$     &  $0.255$ & $0.303$ & $0.419$ & $0.240$ & $0.707$  \\
PSR J1614-2230    &$2.658$     &$1.580$  &  $0.312$   &   $0.277$     &  $0.212$ & $0.282$ & $0.717$ & $0.373$ & $0.997$  \\
PSR J0348+0432    &$2.726$     &$1.571$  &  $0.319$   &   $0.283$     &  $0.204$ & $0.277$ & $0.784$ & $0.400$ & $0.991$ \\[5pt]
PSR J0740+6620    &$3.119$     &$1.727$  &  $0.329$   &   $0.291$     &  $0.192$ & $0.270$ & $0.877$ & $0.435$ & $1.090$ \\
                  &$2.828$     &$1.600$  &  $0.324$   &   $0.286$     &  $0.199$ & $0.274$ & $0.825$ & $0.416$ & $1.010$ \\
\hline
\multicolumn{10}{c}{Gravitational-wave Signals}\\
\hline
GW170817-1        &$2.399$     &$1.588$  &  $0.287$   &   $0.257$     &  $0.240$ & $0.296$ & $0.514$ & $0.285$ & $1.002$  \\
GW170817-2        &$2.304$     &$1.595$  &  $0.278$   &   $0.250$     &  $0.251$ & $0.301$ & $0.440$ & $0.250$ & $1.006$
\end{tabular*}
\end{ruledtabular}}
\end{table}

Next, in Table \ref{Tab:phys_features}, we provide corresponding numerical values of some important physical quantities which identify the physical properties of the pulsar as estimated by QRG model. Obviously, the surface densities are above $\rho_s> \rho_\text{nuc}=2.7\times 10^{14}$ g/cm$^3$ in all cases which is required for a stable quark star, whereas the core densities are few times nuclear saturation density, (2--4)$\rho_\text{nuc}$. Remarkably, the conjectured conformal upper limit on the sound speed is fulfilled for all pulsars even for those which are derived from direct observations. This confirms the role of QRG to obtain this constraint satisfied. This is unlike GR which breaks the conformal constraint in some cases using the same KB matric ansatz \cite{Roupas:2020mvs}. Another interesting feature is the positiveness of the conformal trace anomaly $\Delta$ everywhere inside the pulsar which confirms that the conformal symmetry is broken. On the other hand, it indicates that the stellar matter fulfills the TEC $\mathfrak{T}<0$ everywhere inside the pulsar. The redshift $Z$ is finite although the fluid is assumed to be anisotropic. Finally, we provide the corresponding bag constant $B_{60}=\frac{B}{60~\text{MeV/fm}^3}$ where $B=\frac{1}{4}\rho_s c^2$ and $0.982<B_{60}<1.53$ for physical range. If the latter constraint is fulfilled the strange quark model would be valid in this case. Otherwise, it is not unless some systematic errors are obtained in the data.

According to our calculations as summarized in Table \ref{Tab:phys_features}, the strange quark star model is in tension with NICER measurements of the PSR J0030+0451 radius $R=12.71 \pm 1.19$ \cite{Raaijmakers:2019qny} and $R=13.02 \pm 1.24$ \cite{Miller:2019cac}, since the corresponding bag constants are respectively $B_{60}=0.921$ and $B_{60}=0.911$ below the allowed minimal value. However, the model is in a perfect agreement with the latest analysis of NICER data of the pulsar PSR J0030+0451 \cite{Vinciguerra:2023qxq} which gives $R=11.71^{+0.88}_{-0.83}$ relatively smaller than the previous measurements, and therefore allows for a relatively higher bag constant $B_{60}=1.021$ in the allowed physical range. Also, in the case of PSR J0437-4715, we obtain much lower bag constant value $B_{60}=0.707$, if we consider the relatively much large radius measurement $R=13.6\pm 0.9$ km \cite{Gonzalez-Caniulef:2019wzi}. Therefore, confirmation of the updated NICER analysis of the pulsar PSR J0030+0451 \cite{Vinciguerra:2023qxq}, would validate the QRG scenario of strange quark star model. However, confirmation of the measured radius of PSR J0437-4715 \cite{Gonzalez-Caniulef:2019wzi} would indicate a significant tension with the QRG scenario of strange quark star model.
\newpage


\bibliography{QRastall}

\end{document}